\documentclass{aa} 
\usepackage{natbib}
\usepackage{graphicx}
\usepackage{txfonts}
\usepackage[colorlinks,breaklinks]{hyperref}
\hypersetup{linkcolor=blue,citecolor=blue,filecolor=black,urlcolor=blue}
\usepackage{pdflscape}
\usepackage[T1]{fontenc}
\usepackage{fix-cm}
\usepackage{tablefootnote}

\newcommand{\kms}{km~s$^{-1}$}

\newcommand{\SiII} {\ion{Si}{ii}}
\newcommand{\SiIII} {\ion{Si}{iii}}
\newcommand{\SII} {\ion{S}{ii}}

\newcommand{\CII} {\ion{C}{ii}}
\newcommand{\FeII} {\ion{Fe}{ii}}
\newcommand{\CaII} {\ion{Ca}{ii}}
\newcommand{\SiFeature} {\SiII\ $\lambda6355$}

\begin{document} 

   \title{Exploring the origins of high-velocity features in SNe Ia with the spectral synthesis code \textsc{tardis}.}

    \author{L. Harvey, \inst{\ref{tcd}}
        K. Maguire, \inst{\ref{tcd}}
        A. Holas, \inst{\ref{heidelberg}}
        J.~P. Anderson, \inst{\ref{eso}, \ref{mas}}
	T.-W. Chen, \inst{\ref{taiwan}}
	L. Galbany, \inst{\ref{ice-csic}, \ref{ieec}}
	S. Gonz\'alez-Gait\'an, \inst{\ref{centra}} \\
	M. Gromadzki, \inst{\ref{warsaw}}
	T. E. Müller-Bravo, \inst{\ref{tcd}}
	G. Pignata, \inst{\ref{tarapaca}}
	I. R. Seitenzahl. \inst{\ref{heidelberg}}
        }
    \institute{School of Physics, Trinity College Dublin, College Green, Dublin 2, Ireland \label{tcd}\\
         \email{luharvey@tcd.ie}
         \and Heidelberg Institute for Theoretical Studies, Schloss-Wolfsbrunnenweg 35, 69118 Heidelberg, Germany \label{heidelberg}\
         \and European Southern Observatory, Alonso de C\'ordova 3107, Casilla 19, Santiago, Chile \label{eso}\
         \and Millennium Institute of Astrophysics MAS, Nuncio Monsenor Sotero Sanz 100, Off.104, Providencia, Santiago, Chile \label{mas}\
         \and Graduate Institute of Astronomy, National Central University, 300 Jhongda Road, 32001 Jhongli, Taiwan \label{taiwan}\
         \and Institute of Space Sciences (ICE-CSIC), Campus UAB, Carrer de Can Magrans, s/n, E-08193 Barcelona, Spain \label{ice-csic}\
         \and Institut d'Estudis Espacials de Catalunya (IEEC), 08860 Castelldefels (Barcelona), Spain \label{ieec}\
         \and CENTRA, Instituto Superior T\'ecnico, Universidade de   Lisboa, Av. Rovisco Pais 1, 1049-001 Lisboa, Portugal \label{centra}\
         \and Astronomical Observatory, University of Warsaw, Al. Ujazdowskie 4, 00-478 Warszawa, Poland \label{warsaw}\
         \and Instituto de Alta Investigación, Universidad de Tarapacá, Casilla 7D, Arica, Chile \label{tarapaca}
         }

    \titlerunning{\textsc{tardis} modelling of HVFs in SNe Ia}
    \authorrunning{L. Harvey, et al.}

   \date{Received XXX; accepted YYY}

  \abstract
    {Appearing as secondary higher-velocity absorption components, high-velocity features (HVFs) have been observed in several absorption lines in many Type Ia supernovae (SNe Ia). The frequency and ubiquity of these components in silicon and calcium features specifically indicates that the mechanism through which they form must be common occurrence among the majority of SNe Ia. Here we present modelling of the HVF evolution in a sample of six well observed SNe Ia with the radiative-transfer code \textsc{tardis}. A base model is derived for each of the SNe to reproduce the photospheric velocity components, followed by a grid of simulations with Gaussian enhancements to the density profile at high velocities. We train a set of neural networks to emulate the impact of these density enhancements upon the simulated silicon line profile. These networks are subsequently used within a Markov-Chain Monte Carlo (MCMC) framework to infer the density enhancement parameters that most closely reproduce the HVF evolution. While we obtain good matches for the silicon profile, we find that a single density enhancement alone cannot simultaneously produce the observed silicon and calcium HVF evolution. Our findings indicate that neither the delayed-detonation mechanism, nor the double-detonation mechanism can produce these HVFs, suggesting that something may be missing from the models.}


   \maketitle

\section{Introduction}
\label{intro}
Understood to be the thermonuclear explosions of white dwarfs enabled by some interaction with a binary companion, Type Ia supernovae (SNe Ia) are the most common transients we observe in the night sky. While famed for their use as cosmological distance indicators due to their standardisability via the empirical Phillips relation \citep{phillips_relation_pskovskii, phillips_relation_phillips, tripp98}, extensive observations of these objects in the past few decades have uncovered a large spectrophotometric diversity leading to subdivision into a number of different subclasses (see \citealt{Taubenberger2017_extremes} for a review).

Despite our empirical understanding of thermonuclear SNe, there remain many open questions surrounding their progenitor systems and explosion mechanisms, as well as a number of unexplained observed properties. One such observation is the tendency to find high-velocity features (HVFs) which are secondary, higher-velocity, absorption lines that appear to form in a region of the ejecta detached from the photosphere \citep[e.g.~][]{gerardy2004,mazzali2005,wang_2009,2012fr_spectra1,Maguire_2014}. They are most commonly seen in the \CaII\ and \SiII\ features - the focus of this paper is the \SiFeature\ line - as secondary absorption minima several thousands of \kms\ to the blue of the photospheric features.

Several scenarios have been proposed in the literature to explain these features, including abundance and/or density enhancements in the outer ejecta, as well as changes in the ionisation state of the relevant elements.
Appearing more frequently than in the \SiFeature\ feature, previous studies have mostly focused upon the HVF in the \CaII\ near-infrared (NIR) triplet. Spectropolarimetry of SN~2001el presented by \cite{wang2003} demonstrated high levels of polarisation across the spectrum, with the HV \CaII\ component polarised differently, suggesting that this component is kinematically distinct. A geometrical study of the high-velocity ejecta of this object by \cite{kasen2003} concluded that the HV calcium was produced by a spherically non-symmetric `clumped' shell of material. \cite{thomas2004} studied the HV \CaII\ in SN~2000cx through both its NIR and H\&K absorption features, finding that while 1D models are sufficient to explain the HVF in the NIR triplet, a 3D solution was required to produce the higher velocity structure simultaneously in the \CaII\ H\&K and NIR features. This solution was comprised of a clump of material along the line-of-sight, partially covering the photosphere. The HVF in SN~2003du was proposed by \cite{gerardy2004} to be the result of interaction between the ejecta and the local circumstellar environment. \cite{Mulligan_2017_HVFmodelling} used \textsc{syn++} to investigate if high-velocity \CaII\ NIR features could be explained by the presence of a shell of circumstellar material and put limits on the mass of such a shell. \cite{Tanaka_2006_3DHVF} performed three-dimensional modelling of different geometries of high-velocity material, which was compared to six SNe Ia in  \cite{Tanaka_2008_HVF}. They suggested that asymmetries or abundance enhancements in the outermost layers of the ejecta are likely required to explain the HVFs seen. 

The search for HVFs in early SN Ia spectra by \cite{mazzali2005} found all objects to exhibit HV absorption in the \CaII\ NIR feature, with some objects also displaying these features in the \SiFeature, leading to the conclusion that HVFs are ubiquitous across the SN Ia class. It was suggested that density enhancements instead of abundance enhancements are more likely the explanation for these features given the very high abundance enhancements required to match observations of SN 1999ee \citep{Mazzali_2005_HVF_1999ee,mazzali2005}. \cite{2012fr_velocities} found that \SiFeature\ HVFs were significantly rarer than HVFs in the \CaII\ NIR triplet, and spectra must be obtained at the earliest epochs for the \SiFeature\ HVFs to be detected. High-velocity features of \SiFeature\ and the connection to other observables were also studied in  \cite{Zhao_2015_HVF}. \cite{quimby2006} argued that the ubiquity of these features disfavours line-of-sight effects such as filaments of material. High-cadence spectroscopy of the otherwise normal SN~2009ig by \cite{Marion_2013_2009ig} exhibited HV components to a number of features - \SiII, \SiIII, \SII, \CaII, and \FeII - allowing them to constrain well the transition from HVF to photospheric velocity features (PVF).

The post-maximum light spectra of SN~2012fr were compared to the Chandrasekhar mass deflagration model W7 \citep{Nomoto_1984_W7} using the radiative-transfer code \textsc{phoenix} by \cite{Cain_2018_2012fr}, but the origin of HVFs at early phases was not investigated. Modelling of individual very-early spectra of 14 SNe Ia was performed in \cite{Ogawa_2023_earlyepochTARDIS} using the radiative-transfer code \textsc{tardis} \citep{tardis_original}, but they did not specifically investigate the presence of high-velocity features using distinct density or abundance enhancements above the photosphere. 

In \cite{Harvey_DR2} - hereby referred to as H1 - the Zwicky Transient Facility Cosmology Data Release II (ZTF Cosmo DR2) was searched for \SiFeature\ HVFs, identifying these features in 75 SNe from the 307 that passed our sample cuts. From the phase distribution of the spectra in this sample, we calculated a rate of 78$\pm^{12}_{17}$\%\ per cent of spectra before $-$11~d - with respect to maximum light - to show some HV component to the \SiFeature. When comparing the distributions of various observables between the SNe found to exhibit HVFs and those that did not, we identified no difference between the two populations in terms of SALT light curve width parameter, $x_1$ \citep{Guy2010-SALT2}, peak magnitude in the ZTF\textit{g} band, the magnitude decline rate from peak to 15 d after in the \textit{g} band ($\Delta m_{15,\text{ZTF}g}$), host galaxy stellar mass, or the host galaxy local \textit{g}-\textit{r} colour.

The aim of this paper is to explore the possibility of producing the observed HVF evolution found in a number of SNe Ia by the introduction of a density enhancement in the outer ejecta. We develop models for a number of well-sampled SNe Ia using the radiative-transfer code \textsc{tardis}, aiming simply to reproduce the photospheric features. We then run a number of simulation grids adding in various density enhancements to the base models with the aim of reproducing the observed HVF evolution. This bulk of this work focuses on the \SiFeature, however, HVFs of other species will also be discussed.

In Section~\ref{sec:observations} we describe the collation of our sample and in Section~\ref{sec:method} we present the modelling method. The results of the modelling are analysed in Section~\ref{sec:results} followed by a discussion of the implications of these results in Section~\ref{sec:discussion}.

\section{Observations}
\label{sec:observations}
For our SN Ia sample, for which we will produce custom \textsc{tardis} models, we target events that exhibit clearly separated  \SiFeature\ HVFs  and are photometrically and spectroscopically well sampled within the first two weeks after explosion. We conducted a literature search and found the following suitable events that have early spectra that display distinct \SiFeature\ HVF and PVF: SN~1994D \citep{1994D_analysis,Meikle_1996_1994Dspec,1994D_spectra3}, SN~2009ig \citep{2009ig_spectra1,Marion_2013_2009ig, Chakradhari_2019_2009ig}, SN~2012fr \citep{2012fr_spectra1,Zhang_2014_2012fr,2012fr_dates_photometry}, SN~2018cnw \citep[H1;][]{Rigault_DR2}, SN~2021fxy \citep{2021fxy_spectra2}, and SN~2021aefx \citep{2021aefx_spectra1, 2021aefx_redshift_first_peak}. We consider this modest sample of six SNe Ia with four to five spectra per SN - each SN has a spectrum taken at least 11 d before maximum light. While this sample is biased, consisting of hand-picked objects from the literature, it provides a reasonable number of events with which to study the diversity of strong \SiFeature\ HVFs and their related explosions.

\begin{figure}
 \includegraphics[width = \columnwidth]{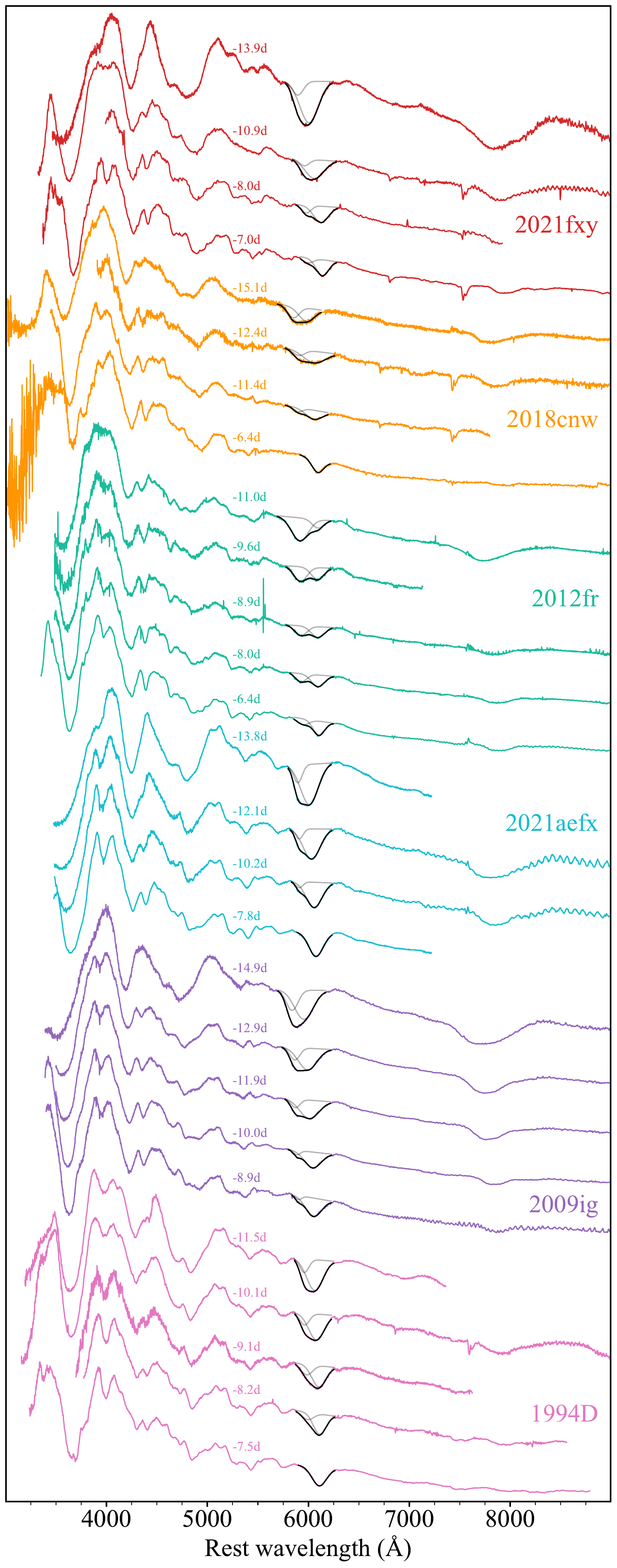}
 \caption{Spectral sequences of the six SNe Ia to be modelled. Fits to the \SiFeature\ feature are displayed in black for each spectrum. In the case of a preferred two-component fit, the individual components are displayed in grey. The spectra are plotted in normalised luminosity and offset, as is the case with all spectra presented throughout this study.}
 \label{fig:spectral_series}
\end{figure}

We collated spectra for each object, covering the full evolution of the HV \SiFeature\ component, aiming for a one day cadence where possible, resulting in 27 spectra for our six SNe Ia. The details of these spectra, including references, can be found in Table~\ref{tab:spectral_log}. We supplemented the first spectrum of SN~2021fxy from \cite{2021fxy_spectra2}, with two new spectra. The first was obtained as part of the ePESSTO+ collaboration \citep{smartt_pessto}, with the ESO Faint Object Spectrograph and Camera v2  \citep[EFOSC2;][]{EFOSC} on the European Southern Observatory's (ESO) New Technology Telescope (NTT). It was reduced and calibrated using the pipeline described in \cite{smartt_pessto}. The second spectrum of SN~2021fxy was obtained with the Spectrograph for the Rapid Acquisition of Transients \citep[SPRAT;][]{SPRAT} on the Liverpool Telescope  \citep[LT;][]{Steele_LT2004}. The SPRAT spectrum was reduced using the pipeline of \cite{Barnsley_FRODOSpec2012} along with a custom Python pipeline \citep[][]{Prentice_2018cow}. While earlier spectra exist for SN~2012fr - as presented in \cite{2012fr_spectra1} - these exhibit sole HV components without PV counterparts (or with very weak PV counterparts) which leads to complications with our modelling method as we are first required to create custom models to fit the PV components. Therefore, the first spectrum we model of SN~2012fr is at $-11$ d from peak. Earlier spectra of SN~2021aefx from \cite{2021aefx_spectra1} were initially included for modelling, however, for these epochs we were unable to produce acceptable PV model fits. This is likely due to the domination from the HV components at very early times.

Each spectrum was flux calibrated to the available photometry, corrected for Milky Way extinction, de-redshifted, and scaled from flux to luminosity. The redshifts, distance moduli, Milky Way extinction values, time of first light and maximum light, as well as the bands used for flux calibration can be found in Table~\ref{appendix:tab:photometry}. Due to the difficulties in estimating host galaxy extinction values, we do not correct for extinction arising in the host galaxies. 

While most of these peak and first light dates were taken from the literature, the date of first-light for both SN~2021fxy and SN~2018cnw were measured through fitting a power law to the early light-curve data. This fitting was performed on ZTF forced photometry data \citep{ztffps}, with the prescription described in \cite{early_lc_fit} and allowing for a variable power law index. The listed phases for SN~2012fr were calculated using the maximum light date from \cite{2012fr_dates_photometry} - as to use the same source for both first light and maximum light dates - and therefore differ slightly from those presented in \cite{2012fr_spectra1}. For SN~1994D, no time of first light could be estimated because neither the date of first light or a measurement of the rise time could be found in the literature. As such, the initial explosion time estimate for the modelling of this object is drawn from the discovery date (MJD 49418) and then pushed backwards accordingly during the modelling. There also exists a discrepancy of one day between the modified Julian date (MJD) of the final SN~2021fxy spectrum used here, drawn from the the public WISeREP archive \citep{WISeREP} and the MJD presented in \cite{2021fxy_spectra2}. With the time since explosion at $\sim$12~d this difference has a negligible effect upon the results presented here. The final corrected spectra are presented in Fig.~\ref{fig:spectral_series}.

We employ the fitting algorithm developed for the HVF search in the ZTF Cosmo DR2 (H1) to fit the \SiFeature\ profiles for these 27 spectra. This algorithm comprised MCMC fitting with single and double component models and selection between the models with the Bayesian Information Criterion. The best fit model in each case can be seen plotted in black in Fig.~\ref{fig:spectral_series}. For each spectrum, we define the continuum regions to the blue and to the red of the \SiFeature\ feature manually, as automated methods may result in incorrect selections for such extreme HVFs as those seen in our sample.

In Fig.~\ref{fig:dr2_comparison} we show how these spectral series compare in terms of the component velocities, the velocity separation between the PVF and HVF ($\Delta$v), and the ratio of the HVF pseudo-equivalent widths (pEW) to the PVF pEW (R$_\text{HVF}$) to the HVF sample found for the ZTF Cosmo DR2. The evolution of the five newly added objects from this paper (SN~2018cnw/ZTF18abauprj was part of the ZTF Cosmo DR2 sample) generally support the findings from H1.

For the pEW ratio, R$_\text{HVF}$, a decrease is seen for individual objects, but no global decrease, matching the results of H1. SN~1994D is an exception; with the smallest velocity separation in our sample, the two components of SN~1994D are highly entangled, leading to a high level of degeneracy between the two components in the spectral  model fits, which is in turn reflected in the measured uncertainties.

H1 saw a global decrease in velocity separation with time, however did not generally observe a decrease in velocity separation for individual objects. Constraints on the evolution of individual objects was limited in the H1 sample as six of the eight multi-epoch objects possessed only two spectra. In this sample we possess fewer targets, however with four to five spectra per target, allowing for tighter constraints upon the phase evolution of HVFs in individual SNe Ia. Here we see an increase in velocity separation with phase in five of the six SNe Ia, implying that the PV component velocity decreases faster than that of the HV component.

As the six SNe Ia explored in this work are well known literature targets with clear HVFs, they tend to have larger velocity separations than the majority of other HVF spectra from the ZTF Cosmo DR2 sample, and therefore, are not representative of the full range of SN Ia HVFs found in nature.

\begin{figure*}
 \includegraphics[width = \linewidth]{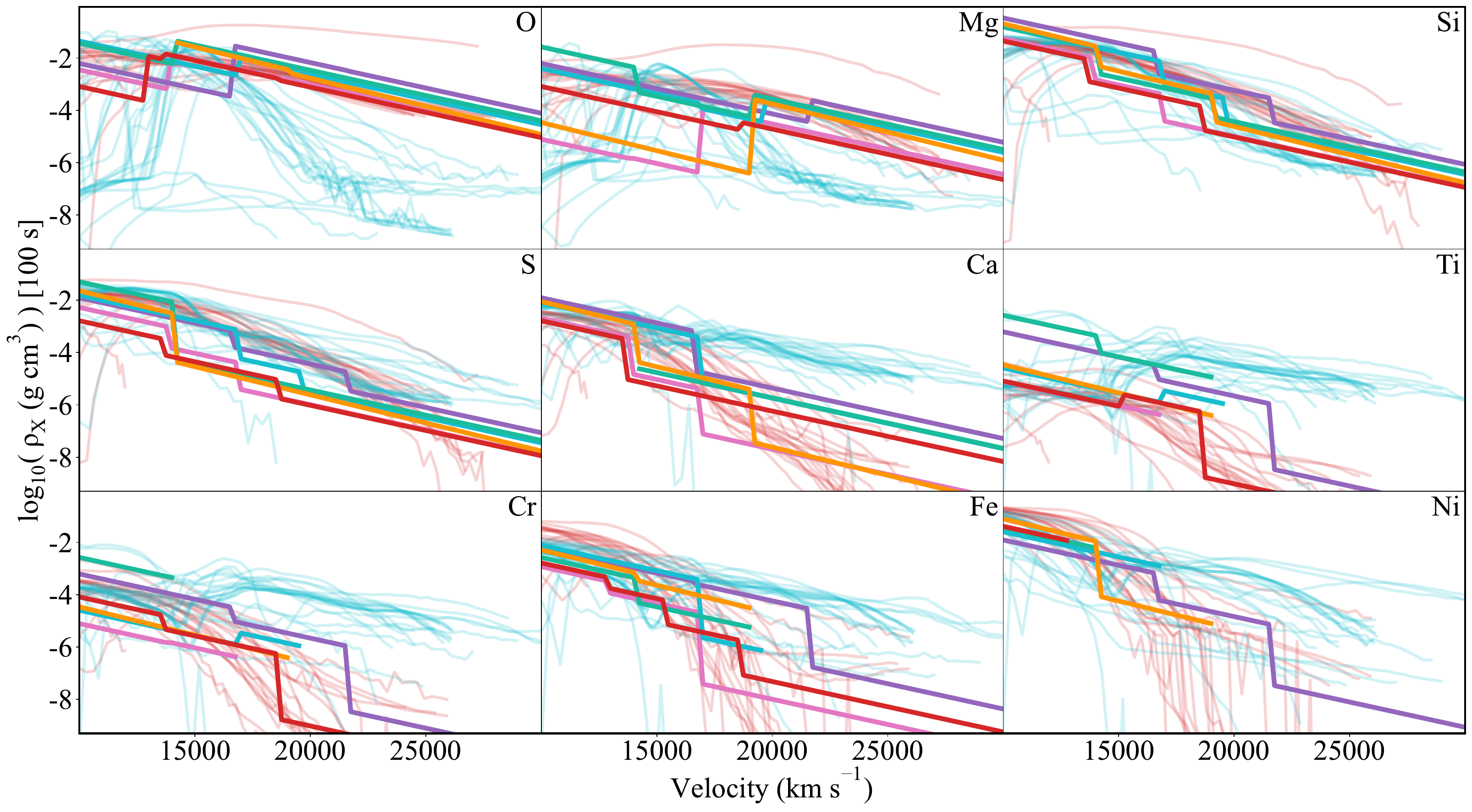}
 \caption{Species densities for the PV models for the six SNe Ia in our sample. These species densities are calculated as the product of the abundance and density profiles and can be used to compare models with different density profiles. The faint red and faint blue profiles correspond to the HESMA delayed-detonation \citep{ddt_models} and double-detonation \citep{double_det_models} models, respectively. The colours of the PV model profiles correspond to the colours in Fig.~\ref{fig:spectral_series}.}
 \label{fig:PV_species_densities}
\end{figure*}

\section{Method}
\label{sec:method}
Our aim in this work was to investigate the origin of high velocity \SiFeature\ features in very early SN Ia spectra. We initially found best matching models for the photospheric components of the spectra and their overall spectral shape, before investigating the impact of high-velocity density enhancements to explain the presence of the high-velocity \SiFeature\ features.
In Section \ref{sec:method:modelling}, we firstly describe our \textsc{tardis} modelling, including our choice of input parameters, such as density and abundance profiles. We then present our methodology for introducing density enhancements in Section \ref{sec:method:modelling:density_enhancements}. In Section \ref{sec:method:neural_networks}, we describe the use of neural network (NN) emulators combined with a Markov-Chain Monte Carlo (MCMC) to determine the \textsc{tardis} parameters for the density enhancements that best match the observed spectra.  

\subsection{\textsc{tardis} photospheric modelling}
\label{sec:method:modelling}
\textsc{tardis} \footnote{\textsc{tardis} version v2024.02.19} is an open-source radiative-transfer code that synthesises spectra from one-dimensional (1D) models of SN ejecta \citep{tardis_original}. The five inputs to a \textsc{tardis} simulation are the time since explosion (\textit{t}), the luminosity (\textit{L}), the inner boundary velocity ($v_{\text{ph}}$), the density profile ($\rho$), and the abundance profile (\textit{A}). The \textsc{tardis} modelling presented here employs the \texttt{nebular} treatment for ionisation, \texttt{dilute-lte} for excitation, \texttt{dilute-blackbody} for radiative rates, and the \texttt{macroatom} treatment (based upon the macro atom described in \citealt{Lucy_macroatom}) for line interaction (see \citealt{tardis_original} for further details). The profiles of the abundance and density are defined as functions of velocity, and divided into discrete shells. The time of explosion for each of the SNe Ia was initially taken as the time of first light (Table~\ref{appendix:tab:photometry}). For each SN, the initial estimate of the explosion time was allowed to be pushed backwards where needed in the development of the photospheric models (PV models) as a control of the temperature to fine tune the match to the observed spectra. This potential difference between the explosion time and the time of first light arises from the time taken for the first photons to escape the ejecta, and is known as the dark phase \citep{Piro2013}. Initial estimates for the inner boundary velocities roughly followed the measurements of the photospheric velocities in the top panel of Fig.~\ref{fig:dr2_comparison}. However, these were varied to improve matches where necessary.

To produce our best matching PV models, we initially remove the \SiFeature\ high-velocity components from our observed spectra by adding the best fitting HV Gaussian (see Section \ref{sec:observations}) to each observed spectrum. While the mechanism forming these HV components likely has significant impacts on other regions of the spectrum, we focus only upon the \SiFeature\ and thus this simplistic treatment is justified. 

\subsubsection{Density profiles}
\label{sec:method:modelling:density_profile}
To develop a custom model for each of the SNe Ia in our sample, we first must choose a density profile. As many of the explosion models have density profiles that are very similar \citep[e.g.~see a comparison in Fig.~10 of][]{2021rhu}, the choice of density profile here is somewhat arbitrary. Therefore, we choose a density profile of the form $10^{mv+d_0}$ with values of $m$ and $d_0$ chosen so that the profile roughly resembles the density profile from the N100 delayed-detonation model \citep{ddt_models} in the region of velocity, $v > 10000$~\kms. Ejecta at velocities of $<$10000~\kms\ are below the photosphere at the epochs considered in this work and are therefore not constrained. The initial density profile as a function of velocity space in the ejecta is chosen for all the objects in this study is:
\begin{equation}
    \rho_0(v) = 10^{mv+d_0}
\label{eqn:base_profile}
\end{equation}
with $m=-1.93\times10^{-4}$ and $d_0=1.43$. This density profile is defined in g cm$^{-3}$ at 100~s post explosion. Throughout the development of the best matching PV model for each SN Ia, tweaks were made to the slope ($m$) and offset ($d_0$) of this density profile to more closely reproduce the evolution of the corresponding \SiFeature\ profile. However, the final density profiles for any SN Ia in our sample did not differ significantly from the original form. 

\subsubsection{Abundance profiles}
\label{sec:method:modelling:photospheric_abundance_profile}
\textsc{tardis} operates with a photospheric approximation, emitting photon packets from a solid inner boundary to pass through the ejected material above. As time progresses with successive spectral observations, this photosphere recedes inwards to lower velocities, revealing previously unseen material. It is this principle that allows for the development of custom abundance profiles with the technique known as abundance tomography \citep{Stehle_abundtomo}. The first spectrum in a series is used to constrain the abundances of the outermost material, which is then kept constant in the modelling of following spectra, with each successive spectrum constraining a new shell of material below the previous photosphere.

The custom models are divided into shells separated by 250~\kms\ with abundances composed of O, Mg, Si, S, Ca, Ti, Cr, Fe, and Ni. C is not included in these models for reasons that shall be discussed in \ref{sec:method:modelling:photospheric_output_models}. We populated the outermost material with elemental abundances that are uniform with velocity. The inner boundary of this layer of material varies from object to object, with a minimum value of 17000~\kms\ for SN~1994D and a maximum of 21500~\kms\ for SN~2009ig. While we did not place strict limits on the allowed abundances in this region, we aimed to follow the relative abundances seen in delayed-detonation models at these higher velocities; this means that it is dominated by unburnt material, with small amounts of intermediate mass elements (IMEs) and little to no iron group elements (IGEs). We initialise the model as 100\% O and gradually introduce other species to fill out the various absorption features.

Each \textsc{tardis} simulation was run with 30 iterations to ensure convergence of the temperature profile, with this value being chosen after it proved to be sufficient in initial manual testing. For these initial grids, the final packet count was set to $10^5$ as this was found to produce spectra with high enough signal-to-noise (S/N) for general comparison to the observed spectra, while keeping the overall computation times reasonable; which is especially important for the $\sim$7000~-~9000 density enhancement simulations per object required in Section~\ref{sec:method:modelling:simulation_grids}.

\subsubsection{Output base photospheric velocity models}
\label{sec:method:modelling:photospheric_output_models}

Through abundance tomography we developed a PV model for each of the six SNe Ia. The synthetic spectra calculated by \textsc{tardis} can be seen in Fig.~\ref{fig:PV_spectral_series}. The goal of each PV model is to closely match the evolution of the Si II PVF, while also matching the overall shape of the spectra, governed principally by the temperature. As we are not focussed upon the relative abundances of the other elements, we are required to designate an element to act as a ‘filler species’ to pad out the normalised abundance in each shell to unity. This element should have minimal impact upon the temperature and opacity of the ejecta, and also must not produce any absorption lines in the region of the feature of interest as to avoid contamination. Therefore, we employ oxygen as this filler species, which results in overly strong O I 7777~Å features in all models. The \CaII\ NIR and H\&K features are not a focus of our PV modelling. Once the best-fitting models are established based on the full \SiFeature\ profile, we will test the plausibility of reproducing the HV calcium features simultaneously with the same density enhancement in the outer ejecta (see Section~\ref{sec:discussion:ca_HVFS}).

The sharp steps seen in the elemental density profiles in Figure~\ref{fig:PV_spectral_series} are not strict constraints of the models, but rather artefacts of the modelling process. Regions of the ejecta were set to be uniform with velocity between the various photospheric velocities for the different spectra. Smoother, continuous curves such as those from the explosion models are more realistic, with our custom models acting as lower resolution approximations.

As the earliest spectroscopic epochs trace the outermost ejecta, they are dominated by contributions from HVFs and it is for these earliest epochs that we find the poorest matches to our PV models. In the earliest epoch of SN~2009ig, \cite{Marion_2013_2009ig} identified HV components in the \FeII\ $\lambda\lambda$5018, 5169 lines, as well as the \SiIII\ $\lambda$4560 line, which causes a shoulder feature to the red of the \CaII\ H\&K. These features can also be seen in the earliest epochs of SN~2021fxy and SN~2021aefx in Fig.~\ref{fig:spectral_series}. 

While the PV model for SN~2021fxy matches well the overall shape of the $-10.8$~d, $-8.0$~d, and $-7.0$~d spectra, the simulated spectrum at the earliest epoch ($-$13.9 d) is overluminous in the blue compared to the observation. We see close agreement in the Si region at $-13.9$~d, $-10.8$~d, and $-8.0$~d, with the simulated PV feature slightly too shallow at the latest epoch.

For SN~2018cnw, while we were able to reproduce the photospheric components in the $-12.7$~d, $-11.8$~d, and $-5.8$~d epochs, we struggled to obtain a good match to the earliest spectrum at $-15.1$~d (the earliest spectrum in our sample) with our PV model. We suspect that at this early phase, the photosphere is at such high velocity that it is in close proximity to the likely density or abundance enhancement causing the HVF and this impacts the PVF strongly. As such, a good match with the PV model alone is not possible.  For completeness we plot the best match we could achieve with the PV model for the $-15.1$~d spectrum along with the good fits to the other epochs in Fig.~\ref{fig:PV_spectral_series}.

As seen in Fig.~\ref{fig:dr2_comparison}, the velocity evolution for SN~2012fr of the observed photospheric component is remarkably flat, and our PV model required a very gradual decrease in the inner velocity parameter to well match this evolution. While this makes for better velocity matches, our PV model produces photospheric \SiFeature\ components that are overly strong in the $-11.0$~d and $-9.6$~d epochs as can be seen in Fig.~\ref{fig:PV_spectral_series}. 

The density enhancements to be introduced to these models in Section \ref{sec:method:modelling:density_enhancements} will bring with them changes to the temperature (and therefore ionisation) structure of the material that may cause strong \CII\ $\lambda$6580 absorption and potentially ruin the match of the model to the observed \SiFeature\ feature. As such we include no carbon in the PV models, instead filling the majority of this outermost layer with oxygen. Once we have obtained a working model for the HVFs, we can introduce a tuned amount of carbon to the outer ejecta, as leftover unburnt material from the original white dwarf.

In Fig.~\ref{fig:PV_species_densities} we present the species density profiles \citep{2021rhu} of the nine elements in our six PV models along with the profiles from the double-detonation  \citep[blue;][]{double_det_models} and delayed-detonation \citep[red;][]{ddt_models} models from the Heidelberg Supernova Model Archive \citep[HESMA;][]{hesma_database}. These species-density profiles are calculated as the product of the mass fraction profile for the species and the density profile of the model, and therefore, can be used to compare models with different density profiles. From Fig.~\ref{fig:PV_species_densities} it is clear to see that the PV models more or less populate the same region of the parameter space as the delayed-detonation models. Slight exceptions to this are the augmented levels of Ti in the models built to match SN~2009ig and SN~2012fr.

\subsection{Density enhancements at high velocity}
\label{sec:method:modelling:density_enhancements}
In this section, we describe how density enhancements were introduced into the best matching PV models obtained in Section \ref{sec:method:modelling} to attempt to match the high-velocity components in the \SiFeature\ feature.
While enhanced abundances of the HVF species (e.g.~Si, Ca) in the upper ejecta have previously been suggested to produce these HV components, there is little justification for abundance enhancements in these regions for individual species. The shell burning region in the double-detonation models can produce this double-peaked abundance profile structure for the IMEs, but it brings with it large amounts of IGEs in the upper ejecta, making for heavily reddened spectra, in disagreement with our sample of observed spectra. This can be seen in Fig.~\ref{fig:PV_species_densities} for the double-detonation models shown in blue, where species densities of IGEs (Ti, Fe, Ni) stay much higher than the delayed-detonation models shown in red at velocities beyond $\sim$18000 \kms. We performed initial testing to explore the possibility of reproducing HVFs through a Gaussian enhancement to the Si abundance relative to the other elements. Even in the regime of extreme abundance enhancements this proved insufficient to reproduce HVFs.

Previous studies have argued that enhancements in the density profile, rather than in the abundance profile, are necessary for HV formation \citep{Mazzali_2005_HVF_1999ee}. During initial tests with density enhancements we observed significantly greater variation in the blue region of the synthesized Si profile than for the abundance enhancement testing, supporting the potential for density enhancements to be responsible for HVF formation as suggested in the literature. These density enhancements alter the temperature and ionisation profiles in such a way that they enable the necessary conditions for HVF formation, highlighting the critical role of density in driving the ionisation balance and spectral features of the outer layers.

The modelling work presented here therefore attempts to reproduce the HVF evolution through Gaussian enhancements to the density profile. These density enhancements are only required in the line-of-sight of the observer and therefore, are not required to be symmetric around the exploding star. Line-of-sight variation of such density enhancements produced by three-dimensional (3D) models may potentially explain the observed diversity found in H1. The choice of Gaussian shaped density enhancements is somewhat arbitrary. Investigation of the effect of the functional form of the density enhancement, such as using alternative profiles like Lorentzian, exponential, or more complex models, would be an interesting direction for future work and could provide deeper insight into the underlying physical processes.

On top of our base PV models we introduce a simple Gaussian density enhancement to the base density profile $\rho_0(v)$, giving a new density profile,
\begin{equation}
    \rho(v) = \rho_0(v) + r\rho_0(b)\exp{\frac{-(v-b)^2}{2c^2}}
\label{eqn:enhanced_profile}
\end{equation}
where $b$ is the velocity location of the centroid of the Gaussian, $c$ is the enhancement width, and $r$ is the amplitude of the enhancement measured as a multiple of the base density at the velocity $b$. Our aim is then to test the impact of various amplitudes, widths, and positions of this density enhancement on the output model spectra. The impact of the injected density enhancements upon the evolution of the \SiFeature\ structure will depend upon the fractional abundance of Si in this high-velocity layer. As such, we also allow this outermost Si abundance to vary within our models while remaining uniform with velocity.

\subsection{Simulation grids}
\label{sec:method:modelling:simulation_grids}
For each of the SNe Ia in our sample we define a grid over which to run \textsc{tardis} simulations with the aforementioned density enhancements. These grids are four-dimensional, corresponding to the three parameters governing the density enhancement, as well as the silicon abundance in the outer ejecta. For each SN our grid spans six values for the amplitude ratio $r$ (0.5, 1, 2, 4, 8, 16), ten widths $c$ between 200 and 2000~\kms\ in increments of 200~\kms, six outer silicon abundances $X_\text{Si}$ from 1\% to 6\% of the total outer layer mass in steps of 1\%, and five velocity values $b$ separated by 1000~\kms, the range of which varies from object to object depending upon the average velocity of the object's HV component. The choice of these values for the Gaussian parameters and Si abundance are somewhat arbitrary but chosen to roughly cover the extremes of the high-velocity \SiII\ components seen in our observed spectra. 

Each spectral series was therefore investigated initially with a grid of 1800 models covering a range of plausible density enhancements. In Section \ref{sec:method:neural_networks}, we will introduce the use of NNs to artificially increase the resolution of the grids of the parameters of the density enhancement, using these 1800 models as the training sample.

\subsection{Neural networks}
\label{sec:method:neural_networks}
Recent years have seen a number of studies implementing machine learning techniques to automate the process of spectroscopic modelling of SNe \citep{Vogl_NN,Chen_AIAI1,Chen_AIAI2,DALEK,OBrien_2002bo,OBrien_91T,Magee_NN}. The general approach is to construct a NN - or a series of NNs - that can emulate the performance of a radiative-transfer code such as \textsc{tardis} in a fraction of a second; a speed increase factor of $\sim$10,000 from the $\sim$5 minutes required for per simulation. This emulator can then be used as a mathematical function in an MCMC framework to calculate the best \textsc{tardis} input parameters. An emulator technique can be employed in this study as a tool to interpolate between the simulation grid points for our density enhancements at high velocity, artificially increasing the resolution of the grid.

As we are solely interested in the formation of the \SiFeature\ PV and HV components, we restrict our investigation to the region enclosed between the two previously defined continuum regions for each spectrum (covering $\sim$5800 - 6200~\AA\ depending upon the spectrum). Each simulation from the grid has some four-dimensional input vector corresponding to the four parameters of the grid ($X_\text{Si}$, $r$, $b$, $c$), which maps to an output spectral vector of length 75, where the region around the \SiFeature\ in the synthetic spectrum has been linearly interpolated at 75 wavelength points. The spectral vector length of 75 was chosen so that the interpolated sections of the spectra would be sampled with wavelength separations of 4--7~\AA, which was shown in H1 to be sufficient for the characterisation of \SiFeature\ lines with HV components. For comparison, the NNs from \cite{DALEK} mapped input vectors of 12 dimensions onto spectral vectors with 500 points covering the range 3400~--~7600~Å.

\begin{figure}
 \includegraphics[width = \linewidth]{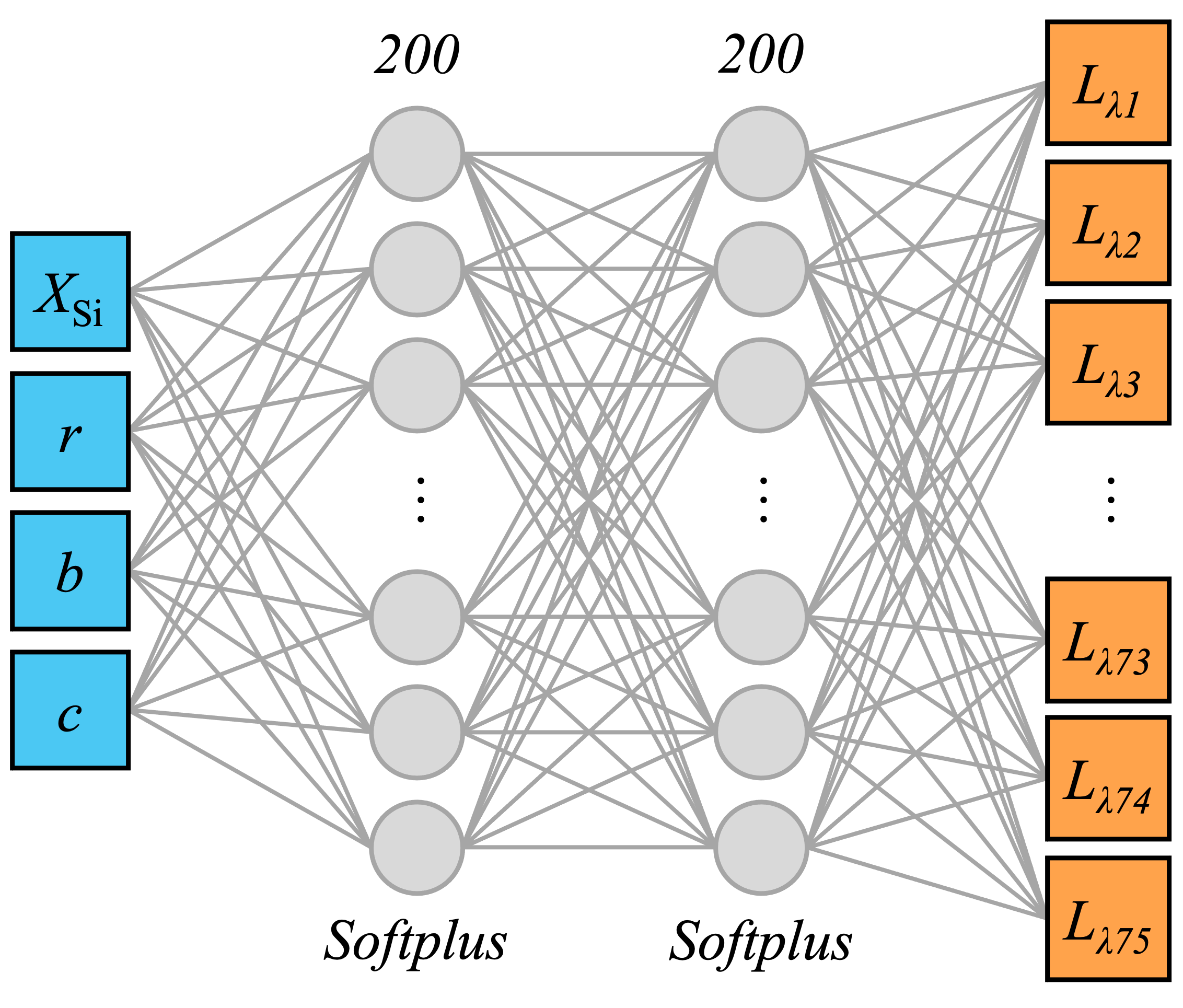}
 \caption{Schematic view of the chosen NN architecture. The blue nodes on the left correspond to the four inputs governing the density enhancement and silicon abundance, which are fed into the input layer. The input layer and hidden layer are both comprised of 200 neurons with a softplus activation function and are represented by the grey nodes. Finally the orange nodes correspond to the normalised luminosity outputs at the 75 wavelength points across the range of the \SiFeature\ feature.}
 \label{fig:nn_architecture}
\end{figure}

\begin{figure*}
 \includegraphics[width = \linewidth]{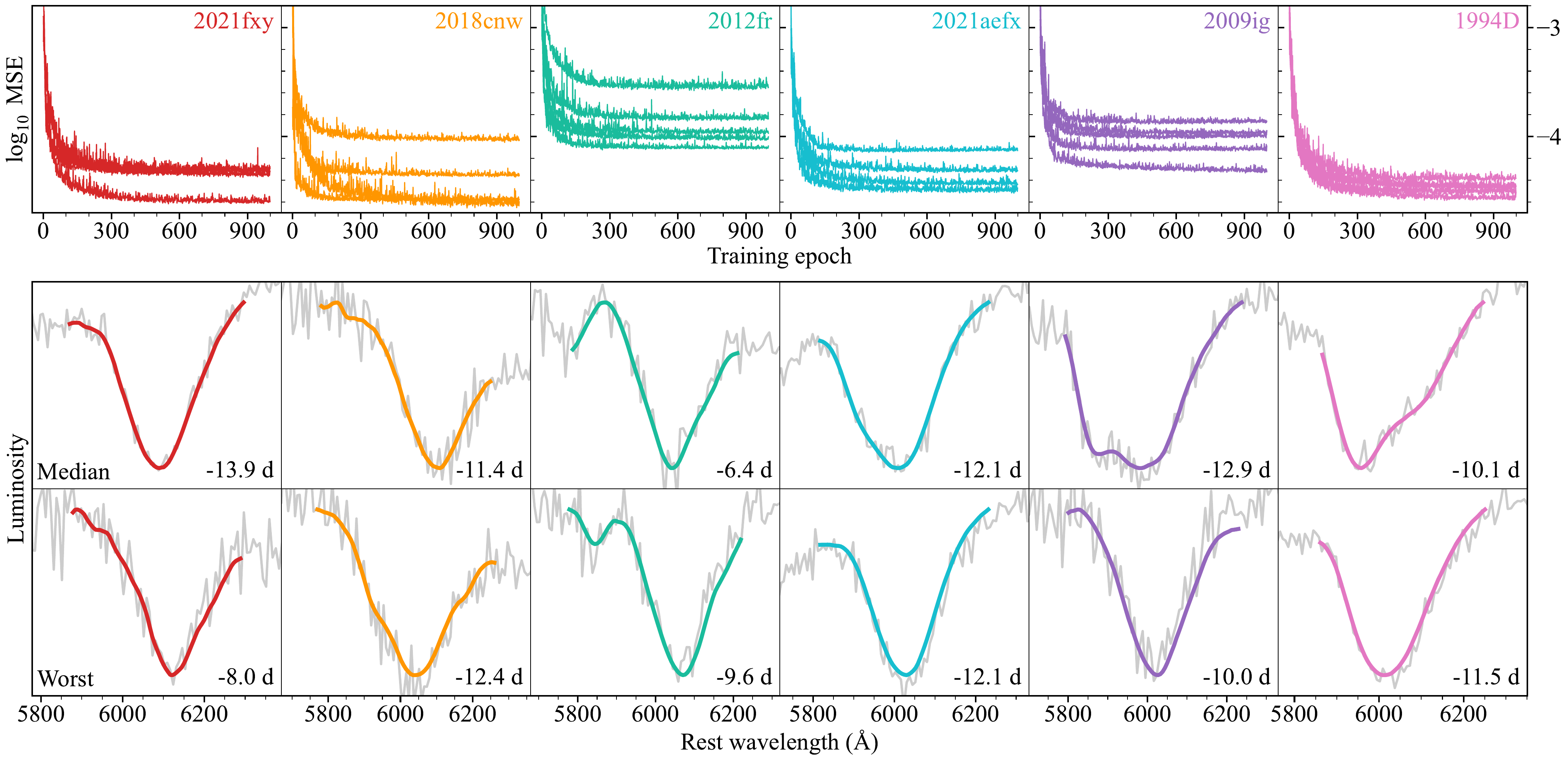}
 \caption{Performance summary of the 27 NNs (four or five per SN based on the number of observed spectral epochs). The evolution of the MSE is shown against the training epoch in the top panels. The middle and bottom panels show the median and worst case predictions, respectively, from the simulated test data corresponding to each SN epoch. The grey lines are the unsmoothed TARDIS outputs and the colours lines are the NN predictions. The median and worst predictions are not necessarily at the same epoch for each SN.}
 \label{fig:nn_performance}
\end{figure*}

\subsubsection{Training}
\label{sec:method:neural_networks:training}
We built and trained our emulator networks with \textsc{tensorflow} \citep{tensorflow} and \textsc{keras} \citep{keras}. Our training sample for each network comprised the full 1800 models of the simulation grid for that SN. While the size of this training set is $\sim$50 times smaller than that of \cite{DALEK}, our input and output dimensions are much smaller (4 versus 12 and 75 versus 500, respectively), and the morphological diversity amongst our simulations is significantly reduced since we are only considering one spectral feature (\SiFeature) compared to a range of 3400 to 7600 \AA\ in their work. We subsequently ran a further 180 simulations per epoch, randomly choosing interpolated input vectors from the simulation grid, which were then split in half to create the validation and test samples.

The validation sample is used throughout the training process to provide feedback on the performance of the network in the form of the mean squared error (MSE) between the predictions and the simulations. While the NNs are not explicitly trained upon the validation data, the resulting MSE evolution curves inform our choices of network hyperparameters and as such these spectra influence the training process indirectly. The test data however have never been seen by the NNs, enabling for an unbiased assessment of the ability of the NNs to accurately emulate the performance of \textsc{tardis}, with different density enhancements post-training.

We perform several preprocessing steps upon the training spectra and input labels before constructing our networks. Being the outputs of Monte Carlo simulations, the generated spectra possess noise which should not be captured by the NNs. As such, the training spectra are smoothed with a Savitzky–Golay filter from the scipy signal module \citep{scipy}, with a window length of 15 and a polynomial order of three. These smoothed spectra are subsequently interpolated to the 75 wavelength points spaced linearly across the \SiFeature\ feature between the predefined continuum regions. We scale the luminosity axis of the smoothed \SiFeature\ feature to have the minimum point at $y$~=~5 and the maximum at $y$~=~10, before finally taking the logarithm with base 10. The input values are mapped to have values between 0 and 1, with these corresponding to the boundary values of the grid. For the parameter $r$ we take the logarithm with base 2 before carrying out this mapping step.

Our initial choice of network architecture corresponds to the best architecture published in \cite{DALEK}. We subsequently tested the impact of changing layer widths and the number of layers, finding no loss of accuracy in reducing the depth of the network to a single hidden layer. Therefore, our chosen architecture for all NNs presented here consisted of two fully connected layers of width 200, with an input dimension of four (corresponding to the density enhancement parameters) and an output dimension of 75 (the emulated spectral vector). The activation function for each of these layers was chosen to be softplus, with the optimizer as nadam. A schematic view of the NN architecture is shown in Fig.~\ref{fig:nn_architecture}. Similarly to \cite{DALEK}, the introduction of a dropout fraction did not improve the accuracy of the emulator predictions and therefore was not implemented. Preliminary testing indicated that the MSE for the validation sample reached a minimum at around $\sim$1000 epochs, after which it gradually tended upwards. The MSE for the training sample however continued to decrease, while at a very slow rate. Such behaviour indicates that after $\sim$1000 epochs we enter a regime of over-fitting, and as such the final networks were trained for a total of 1000 training epochs, with a batch size of 4.

\subsubsection{Performance}
For each spectral epoch, we have trained a NN to predict the impact of the density enhancement on the morphology of the \SiFeature\ feature, therefore amounting to four or five networks per SN depending on the number of spectra for that object. For each of these NNs, we possess a further 90 simulated spectra as our validation set and another 90 spectra as our test set; each corresponding to 5 per cent the size of the training set. The top row of panels in Fig.~\ref{fig:nn_performance} displays the evolution of the MSE of the NN predictions against the validation data with training epoch. In all cases these curves follow the same evolution, with a sharp initial decrease followed by a slow decline to a plateau in the regions of $\sim10^{-4}$.

The MSE curves for networks of SN~2012fr and SN~2009ig plateau at the highest values among the sample. We believe this to be a result of the grid parameter choices for these objects, as they possess the highest range of values for the density enhancement velocity location $b$ (see Section \ref{sec:method:modelling:simulation_grids}). With the enhancement at higher velocities, the formed HVF is more detached from the photosphere and there exists more diversity among the simulation grid in terms of feature shape. This higher morphological diversity still has to be characterised by the NNs with the same number of training spectra as for the less separated features, resulting in lower predicted accuracies. We see the converse effect for the NNs of SN~1994D, with the lowest MSEs as the corresponding grid has the lowest range of $b$ values, reducing the diversity between the models and allowing for more accurate characterisation. Regardless of this difference, the predictions of all networks provide sufficient matches to the test data to proceed with the MCMC fitting.

All of the test set spectra from all the epochs for a given supernova were combined and ranked by the $\chi^2$ difference between prediction and simulation. The median and worst case predictions for each SN are presented in the second and third rows of Fig.~\ref{fig:nn_performance}, respectively. Once again the worst matches of the NN predictions to the simulated models of SN~2012fr and SN~2009ig are the poorest among the six SNe, however still describe the simulated feature reasonably well.

\subsection{MCMC}
\label{sec:method:neural_networks:mcmc}
With the trained NNs reproducing the \textsc{tardis} simulation grids with a high degree of accuracy, we can now run MCMC fits for each of the SNe Ia over all the corresponding observed spectral epochs using \textsc{emcee} \citep{emcee}. We impose uniform priors upon the four input parameters between the boundaries of the grid ($X_\text{Si}$, $r$, $b$, and $c$). For a single spectral epoch, the log-likelihood function takes the form
\begin{equation}
\ln L(X_\text{Si}, r, b, c) = -\frac{1}{2} \sum{ \left(\frac{y_\text{obs}-y_\text{NN}(X_\text{Si}, r, b, c)}{y_\text{obs,err}}\right) ^2 }
\end{equation}
with $y_\text{obs}$ and $y_\text{obs,err}$ as the preprocessed luminosity values of the observed spectra and the corresponding uncertainties respectively, and $y_\text{NN}(X_\text{Si}, r, b, c)$ are the predictions from the relevant NNs. Having set the \textsc{tardis} simulation parameters in the development of the PV models, we have reduced the dimensionality of the parameter space to the four parameters pertaining to density enhancement and the outer silicon abundance. While such assumptions enable probabilistic modelling, there is less flexibility to vary the simulated \SiFeature\ profile and as such perfect matches of the models to the observed spectra are unlikely. To account for model mismatch we assume conservative relative uncertainties on the observed spectra of 5\%. The total log-likelihood for a single SN is calculated as the sum of these individual log-likelihoods for all spectral epochs investigated.

The initial value for all four parameters in all fits was set to be 0.5, i.e. the centre of the parameter space. We ran the MCMC fits with 32 walkers, for an initial burn-in period of 100 iterations, followed by the final chains with 1000 iterations. In four of the six cases these hyperparameter choices resulted in autocorrelation times of $\sim1$ indicating that the chains mixed well and explored the parameter space efficiently. The high autocorrelation times for the remaining two objects (SN~2018cnw and SN~2021fxy) appeared to be caused by the early sections of the chains. The burn-in period for these two objects was therefore increased to 200 iterations, resulting in autocorrelation times of $\sim1$.

As discussed in Section \ref{sec:method:modelling:photospheric_output_models}, the first epoch of SN~2018cnw at $-$15.1 d is not well fit by the PV model. In the MCMC fitting of the HV components of SN~2018cnw, we initially attempted to fit using all four epochs, however due to the poor matching of the $-15.1$~d spectrum this produced a poor match to the \SiFeature\ in all four epochs. Therefore, we repeated the fitting with the first epoch excluded, obtaining a close match to the remaining three spectral epochs.  In the case of the first epoch of SN~2018cnw, the assumption that the density enhancement only affects the formation of the HV feature, and not the overall shape of the spectrum, breaks down and is what we suspect to be the cause of this discrepancy. Simultaneous fitting of the first epoch along with the other epochs would likely require leaving the \textsc{tardis} input parameters as free parameters, which is not computationally feasible. The results of this second fit (excluding the earliest spectrum) are used below.

\section{Results}
\label{sec:results}
\begin{figure}
 \includegraphics[width = \columnwidth]{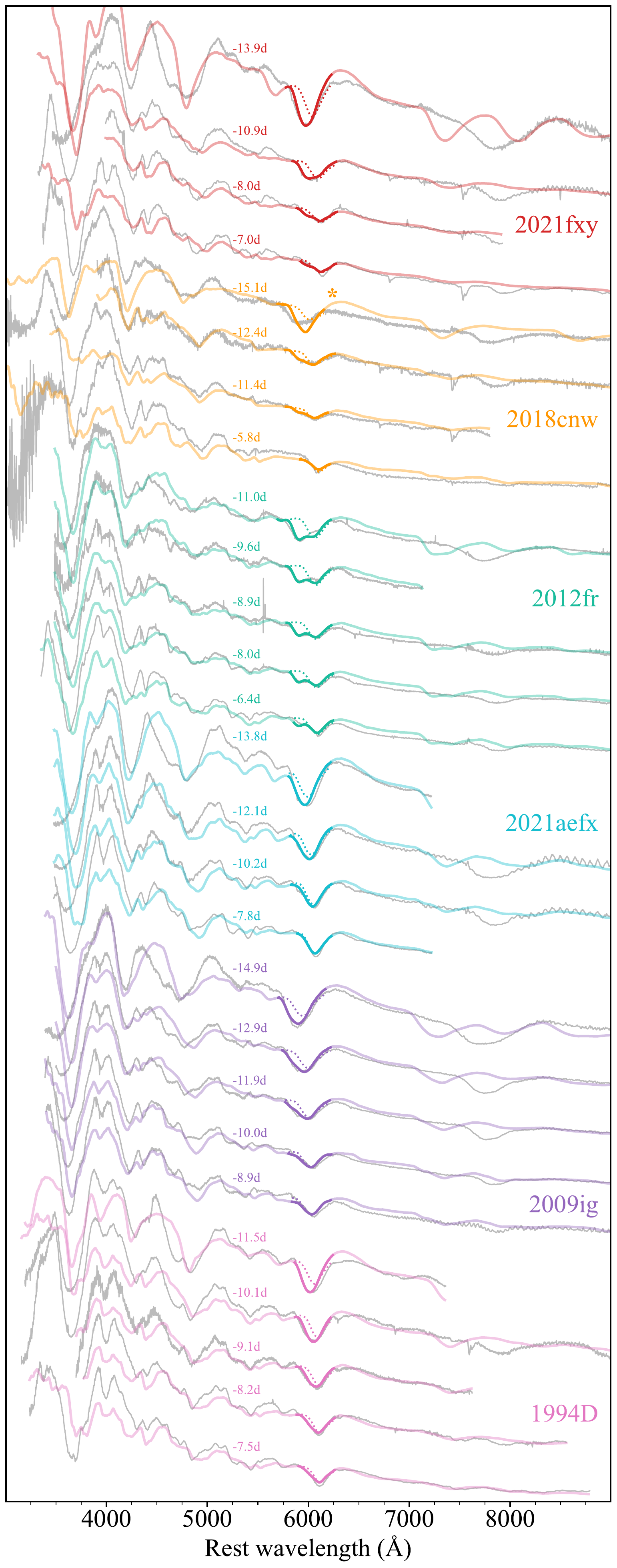}
 \caption{Best fitting density enhancement model spectra (colour solid) compared to observed spectra (grey) and the \SiFeature\ region from the PV simulations (colour dotted). The faint regions of the model spectra correspond to the regions which are not constrained in the MCMC fitting. The asterisk by the $-15.7$~d spectrum of SN~2018cnw indicates that this spectrum was not used in the MCMC parameter inference (see Section \ref{sec:method:neural_networks:mcmc}).}
 \label{fig:spectral_series_DE}
\end{figure}

In this section, we discuss the outputs of our models for the observed \SiFeature\ features in our sample of six SNe Ia. In particular, we are focussed on the contribution from the best fitting high-velocity \SiFeature\ components. As discussed in Section \ref{sec:method:neural_networks:mcmc}, the best-fitting density enhancement parameters are calculated across all spectral epochs simultaneously since we require a single density enhancement to explain the high velocity features and their evolution with time. We present the best NN fit to the \SiFeature\ region of each SNe in Fig.~\ref{fig:spectral_series_DE}. The original best-matching PV only model is also shown for comparison to highlight the contribution of the density enhancement to the feature.

The corner plots for the density enhancement parameters for each of the six MCMCs are shown in Fig.~\ref{fig:corner_plots}. The values of these best fitting Gaussian parameters for the density enhancement and the abundance of \SiII\ at the velocity of the enhancement are given in Table \ref{tab:density_enhancement_parameters}. For SN~2018cnw, SN~2012fr and SN~2009ig, the MCMC contours are well constrained with Gaussian distributions for each of the parameters. However, for SN~2021fxy, SN~2021aefx, and SN~1994D, the MCMC contours are close to the boundaries of the simulation grid and are less well sampled, suggesting that they would prefer values that are higher/lower than the boundaries of the grid.

\begin{figure}
 \includegraphics[width = \columnwidth]{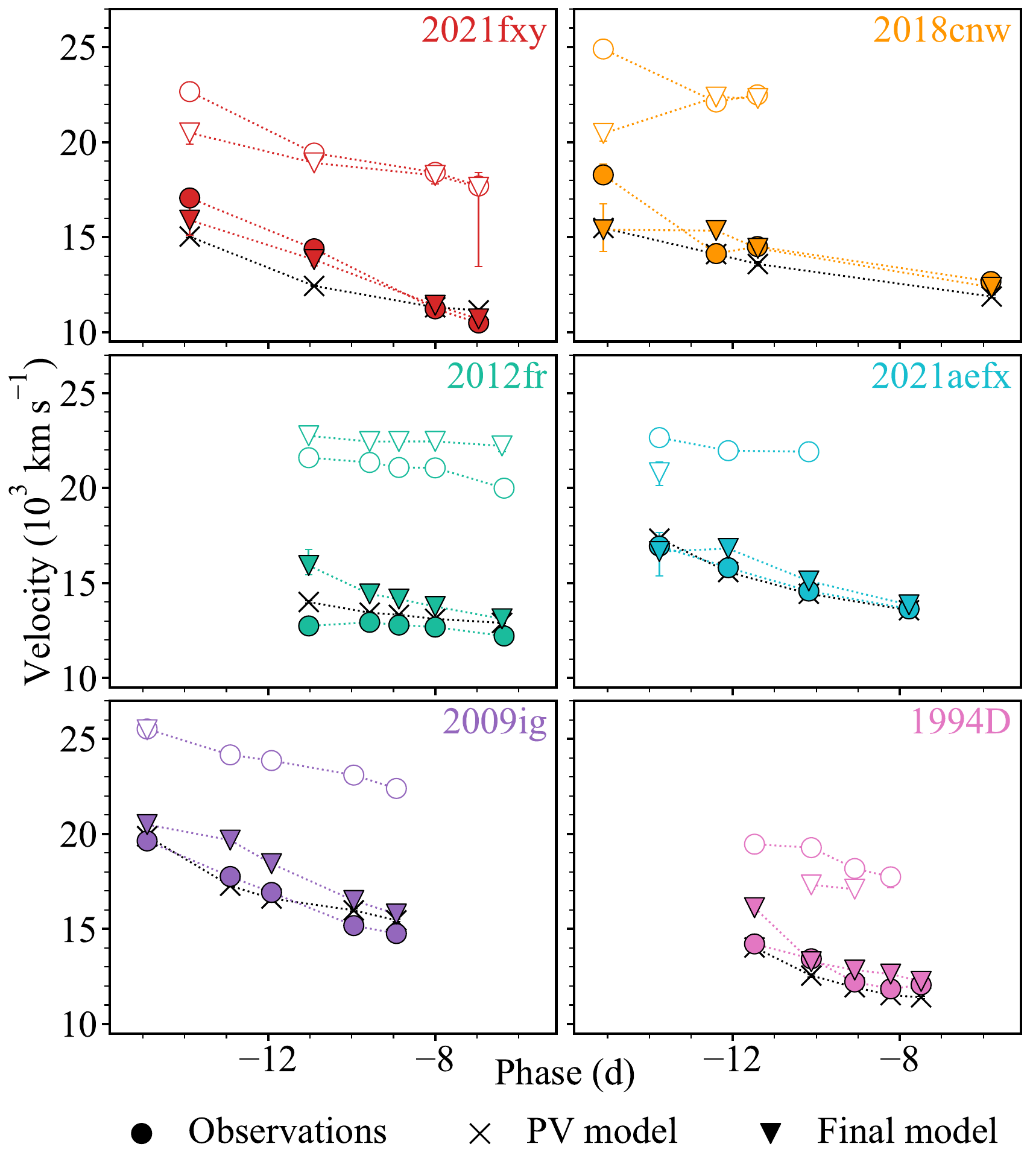}
 \caption{Comparison of the velocities from the best fitting models and the observed spectra. Solid symbols represent the PV components, with open symbols as their HV counterparts. The circles and triangles correspond to the observations and simulations respectively, with the crosses displaying the measurements from the PV model spectra in Fig.~\ref{fig:PV_spectral_series}.}
\label{fig:simulation_velocities}
\end{figure}

A comparison between the HVF and PVF velocities measured for the observed and simulated spectra is shown in Fig.~\ref{fig:simulation_velocities}. These measurements were performed using the code developed in H1. As in H1, we adopt a double-component fit for the simulated features only in cases where this is the preferred model, with $\Delta v$~>~4000~\kms. In general, good agreement is seen suggesting that the models are broadly reproducing the data in the region of the \SiFeature\ feature. In the sections below, we describe the main characteristics of the best fitting models for the \SiFeature\ feature region for each of the six SNe. 

\begin{table}
\setlength{\tabcolsep}{4.5pt}
\caption{Best fitting density enhancement parameters.}
\small
\label{tab:density_enhancement_parameters}
\begin{tabular}{lccccc}
\hline\textbf{Target} & \textbf{\textit{X}$_{\text{Si}}$} (\%) & \textbf{\textit{r}} & \textbf{\textit{b}} (km s$^{-1}$) & \textbf{\textit{c}} (km s$^{-1}$) & \textbf{Mass (M$_\odot$)} \\ 
\hline
2021fxy & 5.9$\pm^{0.1}_{0.2}$ & 1.5$\pm^{0.2}_{0.1}$ & 19500$\pm^{200}_{200}$ & 1900$\pm^{50}_{100}$ & 0.021 \\ [1.6pt]
2018cnw & 4.1$\pm^{0.7}_{0.6}$ & 4.3$\pm^{1.1}_{0.7}$ & 23400$\pm^{300}_{300}$ & 1200$\pm^{400}_{350}$ & 0.019 \\ [1.6pt]
2012fr & 2.2$\pm^{0.3}_{0.3}$ & 6.3$\pm^{0.5}_{0.4}$ & 23100$\pm^{100}_{100}$ & 1750$\pm^{100}_{100}$ & 0.088 \\ [1.6pt]
2021aefx & 4.0$\pm^{0.3}_{0.3}$ & 0.8$\pm^{0.4}_{0.2}$ & 24200$\pm^{400}_{300}$ & 250$\pm^{200}_{50}$ & 0.001 \\ [1.6pt]
2009ig & 3.3$\pm^{0.2}_{0.2}$ & 1.6$\pm^{0.2}_{0.2}$ & 26100$\pm^{200}_{200}$ & 1500$\pm^{150}_{150}$ & 0.013 \\ [1.6pt]
1994D & 5.7$\pm^{0.2}_{0.5}$ & 0.6$\pm^{0.1}_{0.1}$ & 18700$\pm^{400}_{400}$ & 1650$\pm^{250}_{300}$ & 0.010 \\ [1.6pt] \hline
\end{tabular}
\begin{flushleft}
\textbf{Notes.} The mass values correspond to the amount of additional mass introduced to ejecta by the density enhancement, under the assumption that the density enhancement is spherically symmetric. It is important to note that the material from the density enhancements is not required to be new material added to the system, and can be a redistribution of ejecta that would otherwise be located further in at lower velocities. 
 \end{flushleft}
\end{table}

\subsection{SN~2021fxy}
The early spectra and light curves of SN~2021fxy were studied in \cite{2021fxy_spectra2}. They estimated using the blue edge of the \SiFeature\ absorption trough of the first spectrum ($-$13.9 d) that the line-forming region of \SiII\ extends to at least $-$28000 \kms\ but that the \SiFeature\ feature has mostly faded by the $-7.0$~d spectrum.

As seen in Fig.~\ref{fig:spectral_series_DE}, the best fitting density enhancement model provides a good match to the \SiFeature\ in the four spectra of SN~2021fxy. However, as in the PV model alone (see Section \ref{sec:method:modelling:photospheric_output_models}), the PV component strength is still underproduced in the last spectrum at $-7.0$~d. The HV component from the best fitting model matches closely to the observed spectrum in all four epochs.

In Fig.~\ref{fig:corner_plots}, the MCMC contours are seen to overlap the upper boundaries of the grid in the $X_\text{Si}$ and $c$ parameters, indicating that a more extended density enhancement combined with a higher Si abundance would be marginally preferred for SN~2021fxy. The width of the density enhancement for SN~2021fxy is already the highest for our sample and the Si abundance is the joint largest with SN~1994D. The velocities measured using Gaussian fits (as described in Section \ref{sec:observations}) to the best fitting models and the observed spectra are in excellent agreement for SN~2021fxy, again suggesting that the model is a good match to the data (Fig.~\ref{fig:simulation_velocities}). 

\subsection{SN~2018cnw}
\label{sec:results:18cnw}
Identified as having a HV component in H1 in three of the four available epochs, SN 2018cnw exhibits the largest velocity separation measured in the ZTF Cosmology DR2. With four epochs from $-$15.1 to $-$5.8 d covering $\sim$9 d -- the largest phase range in this sample -- the HV component is seen to fade away completely to leave the solitary photospheric component.

As discussed in Section \ref{sec:method:neural_networks:mcmc}, the earliest spectrum of SN~2018cnw was excluded from the MCMC fitting. As expected there is therefore a significant difference seen in the velocities obtained from the Gaussian fits to the observed and best fitting models for this first epoch in Fig.~\ref{fig:simulation_velocities}. However, the velocities measured from the following epochs are in good agreement with the fits to the observed data. The reproduction of the \SiFeature\ structure in these later three epochs shown in Fig.~\ref{fig:spectral_series_DE} are seen to be good matches and the MCMC contours are also well constrained (Fig.~\ref{fig:corner_plots}).  

\subsection{SN~2012fr}
With the largest velocity separation in the sample, and higher than any object found in the ZTF DR2 (H1), SN~2012fr represents the extreme end of the HVF population. An extensive observational study of the HV features of \SiII\ and \CaII\ was first presented in \cite{2012fr_spectra1}. The distinct nature of the HVF for SN~2012fr was also confirmed in other studies \citep{Zhang_2014_2012fr,2012fr_dates_photometry}. For the spectra investigated in this work, we measure the HV component to have a mean separation of 8300~\kms\ from its PV counterpart, making for a double component profile that can be easily identified by eye. Our spectral series for SN~2012fr spans $\sim$5 days from $-11$ to $-$6 days with respect to peak. These spectra demonstrate the full evolution of a HVF, from starting off the dominant component, rapidly weakening to be of equal strength as the PV component, and then to fading away completely.  We find consistent \SiFeature\ HV and PV component velocities compared to the values of \cite{2012fr_spectra1} of $\sim$22,000 to 20,000 \kms\ for the HVF and $\sim$13,000 to 12,000 \kms for the PVF, over the same phase range as studied here.

\begin{figure}
 \includegraphics[width = \columnwidth]{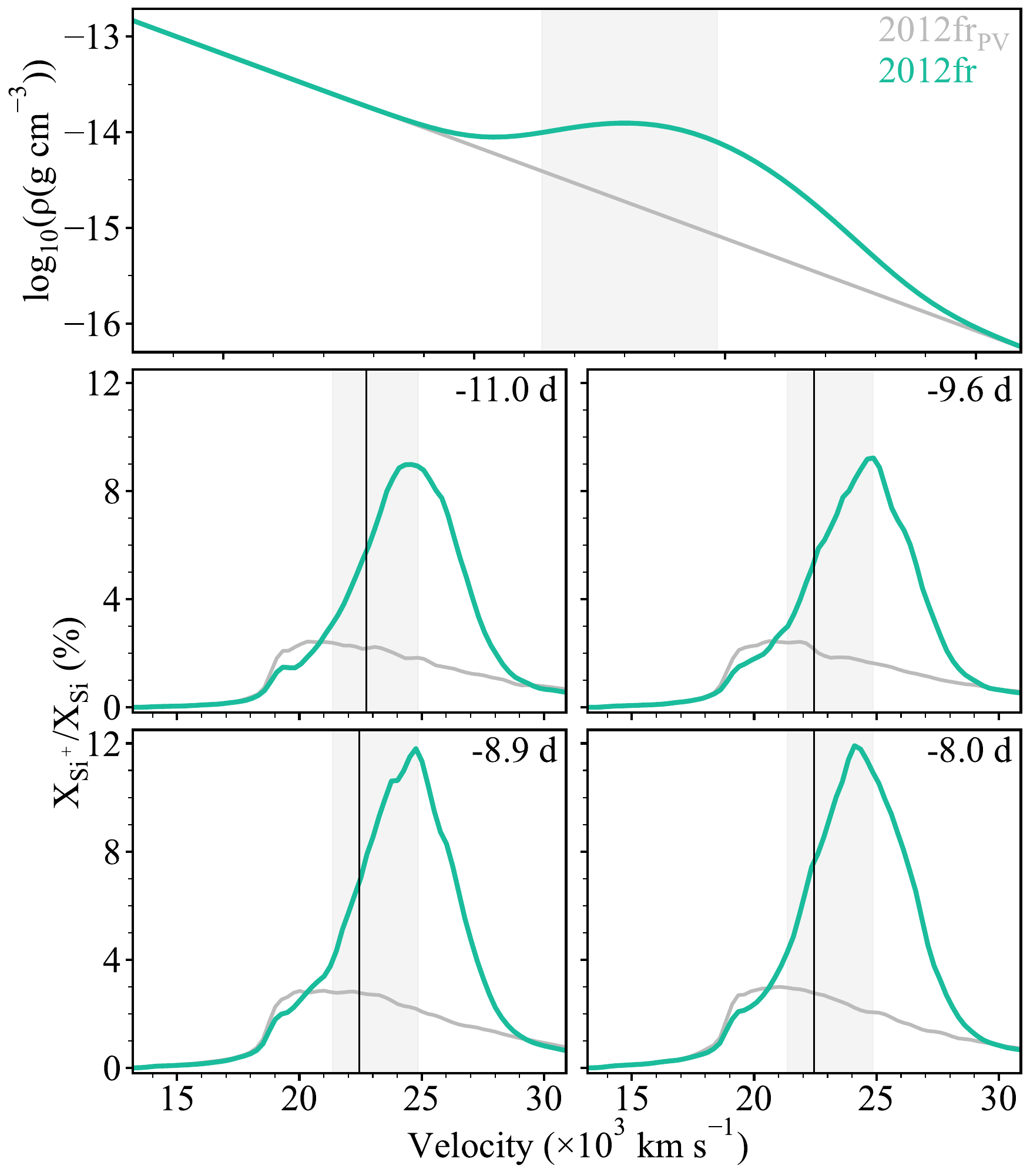}
 \caption{\textit{Top:} The density profiles from the PV model (grey) and the density enhancement model (green) for SN~2012fr. \textit{Bottom:} The relative fraction of silicon material in the singly ionised state for the PV model (grey) and the density enhancement model (green) across the first four epochs of SN~2012fr. The black vertical lines correspond to the measured HVF velocities from the synthesised spectra. The shaded region represents the velocity range centred at $b$ with width $2c$ (i.e., from $b - c$ to $b + c$), where $b$ and $c$ are the centroid and width of the Gaussian density enhancement, respectively, as given in Table~\ref{tab:density_enhancement_parameters}.}
\label{fig:ionisation_profiles}
\end{figure}

In Fig.~\ref{fig:ionisation_profiles}, we show the fractional abundance of singly ionised silicon (Si$^+$) for both the PV model and the density enhancement model across multiple epochs. The introduction of the density enhancement leads to elevated temperatures throughout the ejecta, up to and including, the region of the enhancement. These higher temperatures suppress recombination, resulting in lower Si$^+$ abundances around $\sim$20,000~\kms, just below the core of the density enhancement. At higher velocities, within the enhancement itself, the increased electron density drives a sharp rise in the recombination rate, leading to significantly higher Si$^+$ abundances in the density enhancement model. This region of enhanced Si$^+$ is responsible for producing the high-velocity features (HVFs) observed in the synthetic spectra.

In our best fitting density enhancement model, we see the velocity of the PV component pushed slightly higher compared to that in the PV model, resulting in the small offset we see in Fig.~\ref{fig:simulation_velocities}, where the PV velocities measured from the best fitting models are slightly higher than the observed spectra. We similarly see a velocity difference between the observed and simulated HV components.

Our best fitting model with a density enhancement for 2012fr closely matches the observed HVF evolution of the object (Fig.~\ref{fig:spectral_series_DE}), and consists of a low outer abundance of Si (the lowest in our sample) with a strong (highest in our sample) and wide density enhancement at a large separation from the photosphere. This result implies that the highly separated HVFs find their origins in fairly extreme density enhancements, which is reflected in the fact that they are so rare.

Being such an extreme object, we also find the strongest parameter correlations in the fitting parameters posteriors. As seen in the corresponding panels of Fig~\ref{fig:corner_plots}, we find the outer silicon abundance ($X_\text{Si}$) to correlate positively with the enhancement centroid velocity ($b$) and negatively with the enhancement strength ($r$). As the density enhancement is moved out to higher velocities where there is less material, we require a higher concentration of silicon to form the HVF. Conversely, as the concentration of silicon increases, we require the density enhancements to be less extreme in amplitude.

\subsection{SN~2021aefx}
SN~2021aefx has been the focus of a number of studies at both early \citep{2021aefx_photometry,2021aefx_spectra1, 2021aefx_redshift_first_peak} and late times, including extensively with James Webb Space Telescope (JWST), in both imaging  \citep{Chen_2023_2021aefx_lateJWSTimaging} and spectroscopy \citep{Kwok_2023_2021aefx,DerKacy_2023_2021aefx_lateJWSTMIRIspec,Blondin_2023_2021aefx_modelling,Ashall_2024_2021aefx_lateJWSTspec}. A prominent HVF of the \SiFeature\ was identified in the earliest spectra and the feature was seen to evolve rapidly with time. 

The posterior contours overlap the lower grid boundary for the width of the Gaussian density enhancement parameter $c$, resulting in a density enhancement confined to a very narrow range of velocities; significantly narrower than the other events. While the best density enhancement model fills out the two component \SiFeature\ feature well (Fig.~\ref{fig:spectral_series_DE}), the specific shape of the absorption structure is not well matched, with the MCMC fit tending towards a more entangled PV/HV pair that resembles more of a single broad component, compared to the fairly distinct components seen in Fig.~\ref{fig:spectral_series}. This results in a two component fit being preferred in only the first epoch for the density enhancement model, as opposed to the first three epochs for the observations. This can be seen in the SN~2021aefx panel of Fig.~\ref{fig:simulation_velocities} for which we only possess a HVF velocity measurement at $-13.8$~d.

\subsection{SN~2009ig}
We measured the highest HVF velocities in our sample for SN~2009ig, in agreement with high velocities measured in previous studies of this SN \citep{2009ig_spectra1, Marion_2013_2009ig, Chakradhari_2019_2009ig}. However, as the PV component velocities are also at the higher end of the range, we do not see as extreme a velocity separation as we saw for SN~2012fr (Fig~\ref{fig:dr2_comparison}). Due to these high velocities, the best density enhancement model identified by the MCMC fitting has the highest density enhancement velocity ($b$) among all the objects investigated.

While the first epoch of SN~2009ig is at a similar phase to the first epoch of SN~2018cnw and has a similar inner boundary velocity, we did not encounter the same issues with regards to matching the \SiFeature\ velocity in the PV model, nor did we require the exclusion of the first spectral epoch in the MCMC fitting. This is likely due to the density enhancement being at such high velocities, causing less disruption to the rest of the spectrum. Although the density enhancement fills out the velocity extent of the feature, the fitting of the simulated features with the HVF classification code prefers a double-component classification only in the earliest epoch. This can be seen in Fig.~\ref{fig:simulation_velocities}. The single-component classification for the simulated features at the remaining four epochs indicates that the simulated line profiles closely resemble single broad Gaussians.

\subsection{SN~1994D}
\cite{1994D_spectra3} considered the \SiFeature\ of SN~1994D as a single component in their spectra but noted that there was a very steep decline in the velocities across the earliest spectra ($-$11 to $-$8 d from peak), which we now interpret as the phases where an additional component from a HVF is present. SN~1994D exhibits the smallest velocity separation in our sample, appearing to bridge the gap between broad-line SNe Ia and those with clearly resolved HV components. With a mean HVF velocity of only $\sim18000$~\kms\ over the epochs investigated, we extended the uniform abundance high-velocity shell in our models inwards to 17000~\kms. The velocity range for the density enhancement chosen for the simulation grid ran from 18000 -- 22000~\kms.
 
While the best-fitting density enhancement model matches well the evolution of the SN~1994D \SiFeature\ feature in Fig.~\ref{fig:spectral_series_DE}, the MCMC results cluster up against the boundaries of the grid. The parameter contours in Fig.~\ref{fig:corner_plots} overlap the upper limit on the outer silicon abundance ($X_\text{Si}$), as well as the lower limit on the density enhancement strength ($r$). In contrast to the extreme density enhancements required for the highly separated HVFs, higher silicon abundances with very slight density enhancements appear to give rise to the less separated HVFs.

These small density enhancements are of the same scale as the general variation we see in the angle-averaged density profiles from the HESMA simulations, and are therefore likely to be very common. The regularity of such density enhancements agrees with the fact that these smaller velocity separations were found to be the most common in H1.

In fitting the simulated \SiFeature\ features, a single-component model was preferred at the earliest epoch, with the simulated line profile more Gaussian in shape than in the observed spectrum. In the two following epochs the classification code prefers a double-component model, as seen for the observed spectra.

\begin{figure}
 \includegraphics[width = \columnwidth]{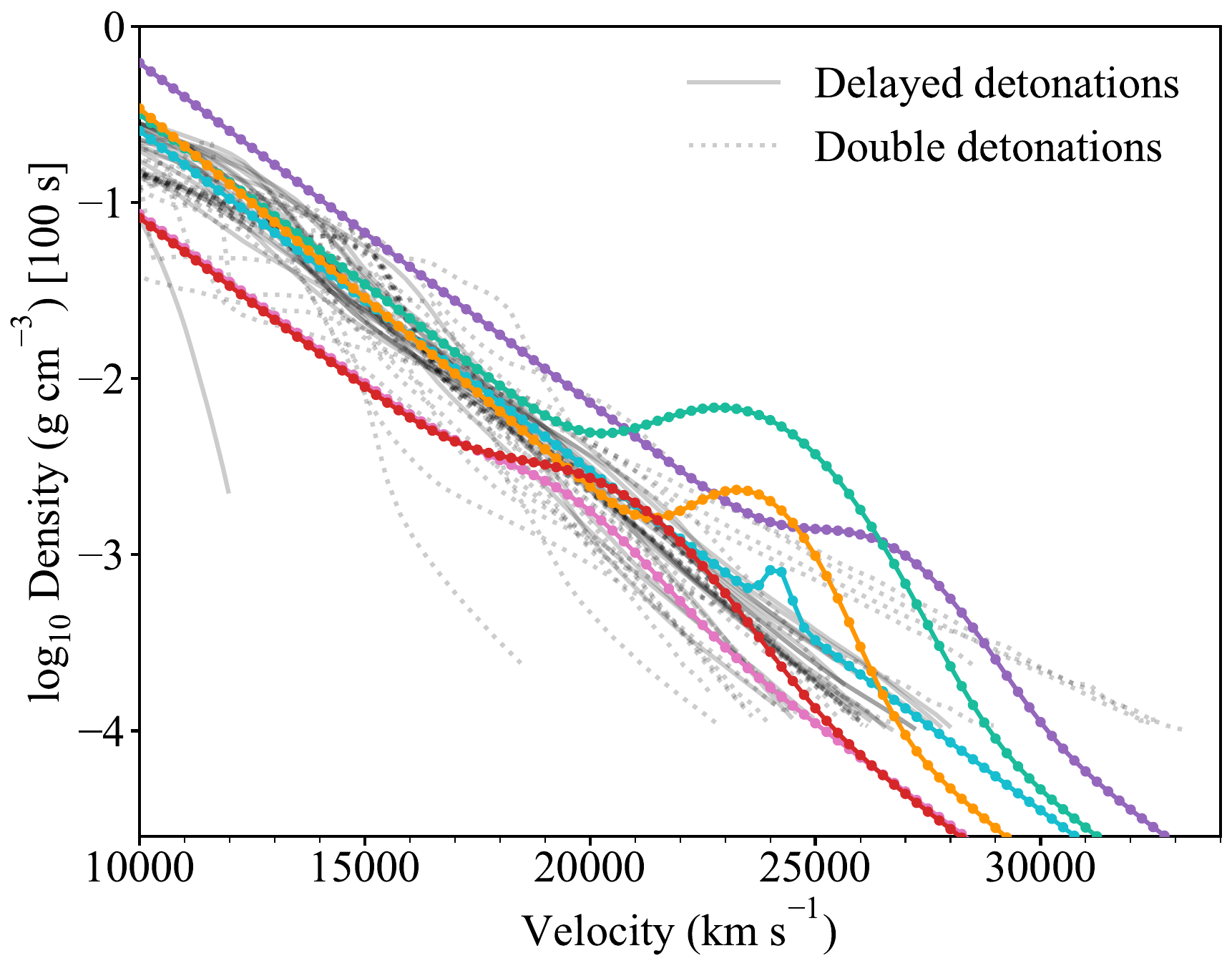}
 \caption{The best fitting enhanced density profiles for each of the six supernovae. The PV model density profiles for the targets are shown as the coloured dotted lines, with the gray solid and dashed lines corresponding to the angle-averaged density profiles from the delayed-detonation and double-detonation simulations from HESMA respectively.}
\label{fig:enhanced_density_profiles}
\end{figure}

\section{Discussion}
\label{sec:discussion}

In this section, we discuss the combined results of our \textsc{tardis} modelling of six SNe Ia showing high-velocity components in their \SiFeature\ features in very early time spectra ($-$14 to $-$6 d with respect to maximum light). In Section \ref{sec:discussion:origin_HVSiII}, we discuss the properties of the density enhancements required in our study to explain the diversity of the observed \SiFeature\ HV features. In Section \ref{sec:discussion:ca_HVFS}, we alter the outer Ca abundance of our best fitting models to test whether the Ca HVFs can be simultaneously formed by the same density enhancement and in Section \ref{sec:discussion:hydro}, we discuss our results in the context of the hydrodynamical delayed-detonation \citep{ddt_models} and double-detonation models of \cite{double_det_models}. 

\subsection{Density enhancement morphology and origin}
\label{sec:discussion:origin_HVSiII}
In Fig.~\ref{fig:enhanced_density_profiles}, we show the density profiles for each of our six SNe with the density enhancement that best matches the observed HV \SiII\ features. There is seen to be significant variation in the positions of the density enhancements, their widths, and their strengths. As shown in Table \ref{tab:density_enhancement_parameters}, the centroid of the enhancement varies from $\sim$18,700~\kms\ for SN~1994D to 26,100~\kms\ for SN~2009ig. The centroid of the enhancement is seen to broadly correlate with the measured HV component velocities shown in Fig.~\ref{fig:dr2_comparison}, where SN~1994D had the lowest HV component velocities and SN~2009ig had the highest. 

The strength of the density enhancement also varies between the SNe Ia in the sample, with the strongest density enhancements typically required to produce the more extreme velocity separations. As by definition the highly detached HV components are formed far above the photosphere, they have densities far lower than the photospheric density and as such require these fairly extreme enhancements to produce line forming regions. The converse affect can be seen for SN~1994D where the HV line forming region has densities much closer to that of the photospheric region and as a result only a slight density enhancement is required. The stronger density enhancements are likely far less prevalent in nature and therefore, explain the scarcity of events such as SN~2012fr.

These density enhancements constitute additional material introduced atop the base density profile, and a mechanism is therefore required to provide this additional mass. Firstly, we consider the extreme-velocity material (above the density enhancement region) within the supernovae and whether these enhancements may be the bunching up of this faster moving material, relocating it to slower velocities. While potentially plausible for the weakest undulations - as seen in SN~1994D - the majority of the enhancements found here invoke masses significantly larger than the masses found at velocities above the density enhancement in literature models. This implies that these enhancements are not the result of the compacting of faster moving material.

Another potential candidate for the source of this additional mass is circumstellar material (CSM) surrounding the system. This material however moves at non-explosive velocities and we would therefore require some mechanism to accelerate the CSM to the velocities of the density enhancements ($\sim$20,000--30,000~\kms). CSM is therefore also unlikely to provide this additional mass.

The most promising candidate for this additional mass is likely to be the supernova itself. Whether linked to a particular explosion mechanism, or produced through some asymmetry of the progenitor system, differences in the distribution of kinetic energy from the explosion could impact the distribution of material at high velocities. We therefore propose these density enhancements to be generated by the supernova explosions themselves.

\begin{figure}
 \includegraphics[width = \columnwidth]{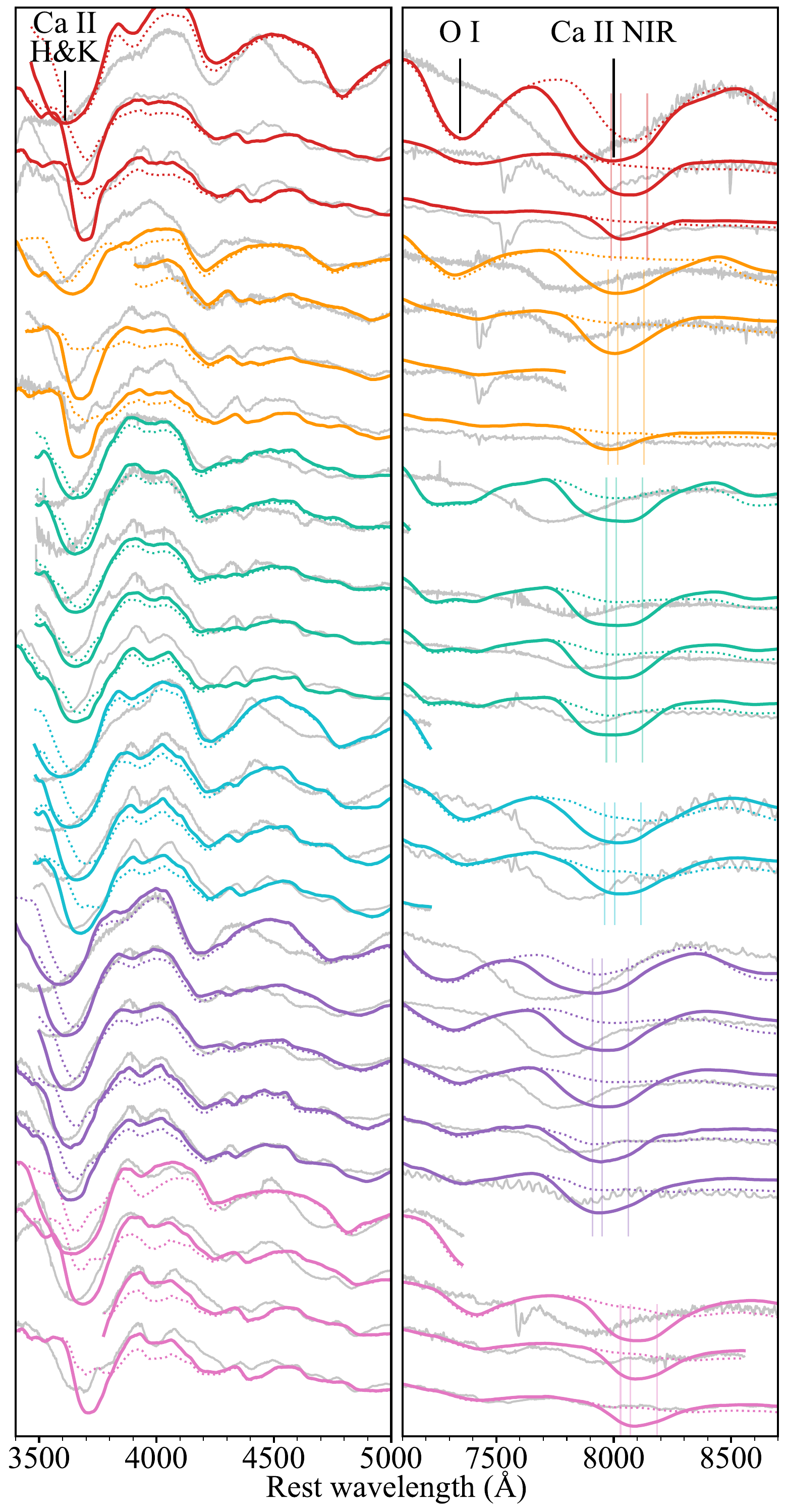}
 \caption{The \CaII~H\&K and NIR features synthesised by the best-fitting density enhancement profiles with Ca abundances as defined in the PV models (dotted) and with 5\% outer calcium (solid). The coloured vertical lines correspond to the NIR wavelengths at the inner velocity boundary of the outer ejecta shell.}
\label{fig:ca_abundances}
\end{figure}

\subsection{Calcium HVFs}
\label{sec:discussion:ca_HVFS}
With best-fitting density enhancement models for the \SiFeature\ HVFs determined, we turn our focus to the HV components of the \CaII\ H\&K and NIR features. To test if the current enhanced density profiles can simultaneously reproduce the \SiFeature\ and \CaII\ HVFs, we ran a small grid of 20 models for each SN Ia varying the outer calcium abundance - in the same region as for the outer silicon abundance - over the range 0.001-5\% with a logarithmic spacing. Increasing the calcium abundance yielded stronger \CaII\ H\&K and NIR features, however the \CaII\ features formed at velocities far lower than the observed \CaII\ HV components. The synthetic spectra from the models with highest calcium abundances (5\%) can be seen in Fig.~\ref{fig:ca_abundances} over the H\&K and NIR regions for all epochs with wavelength coverage in at least one of these regions in the corresponding observed spectrum.

As is clear from Fig.~\ref{fig:ca_abundances}, we get strong \CaII\ NIR formation at the inner edge of this high-velocity region in which we increased the Ca abundance - indicated by the vertical coloured lines. In our observed spectra we do not find \CaII\ NIR components with velocities this low and therefore, require significantly less calcium around the lower boundary of this region, while retaining the silicon abundance constrained by the MCMC fitting. In some of the earliest epochs in the simulated spectra, we also see some higher velocity \CaII\ NIR absorption coming from the density enhancement region. However, these sit thousands of \kms\ lower than the observed \CaII\ NIR HV components.  Therefore, we require low Ca abundances in the vicinity of the \SiFeature\ density enhancement and suggest that the observed \CaII\ NIR HVFs form from a secondary density enhancement higher up in the ejecta.

From this we speculate on the potential for three line-forming regions: the photosphere, the density enhancement for the \SiFeature\ HVFs (Si-dominated density enhancement), and another higher velocity density enhancement for the \CaII\ HVFs (Ca-dominated density enhancement). As we have never observed any triple \SiFeature\ components - with the third component aligned with the \CaII\ HVF - we postulate that there is a drop off in the silicon abundance in the outermost ejecta where we find this density enhancement. We therefore would expect the ratio of Ca/Si to decrease as we move from high velocities into the lower velocity material where the Si becomes more dominant.

\begin{figure}
 \includegraphics[width = \linewidth]{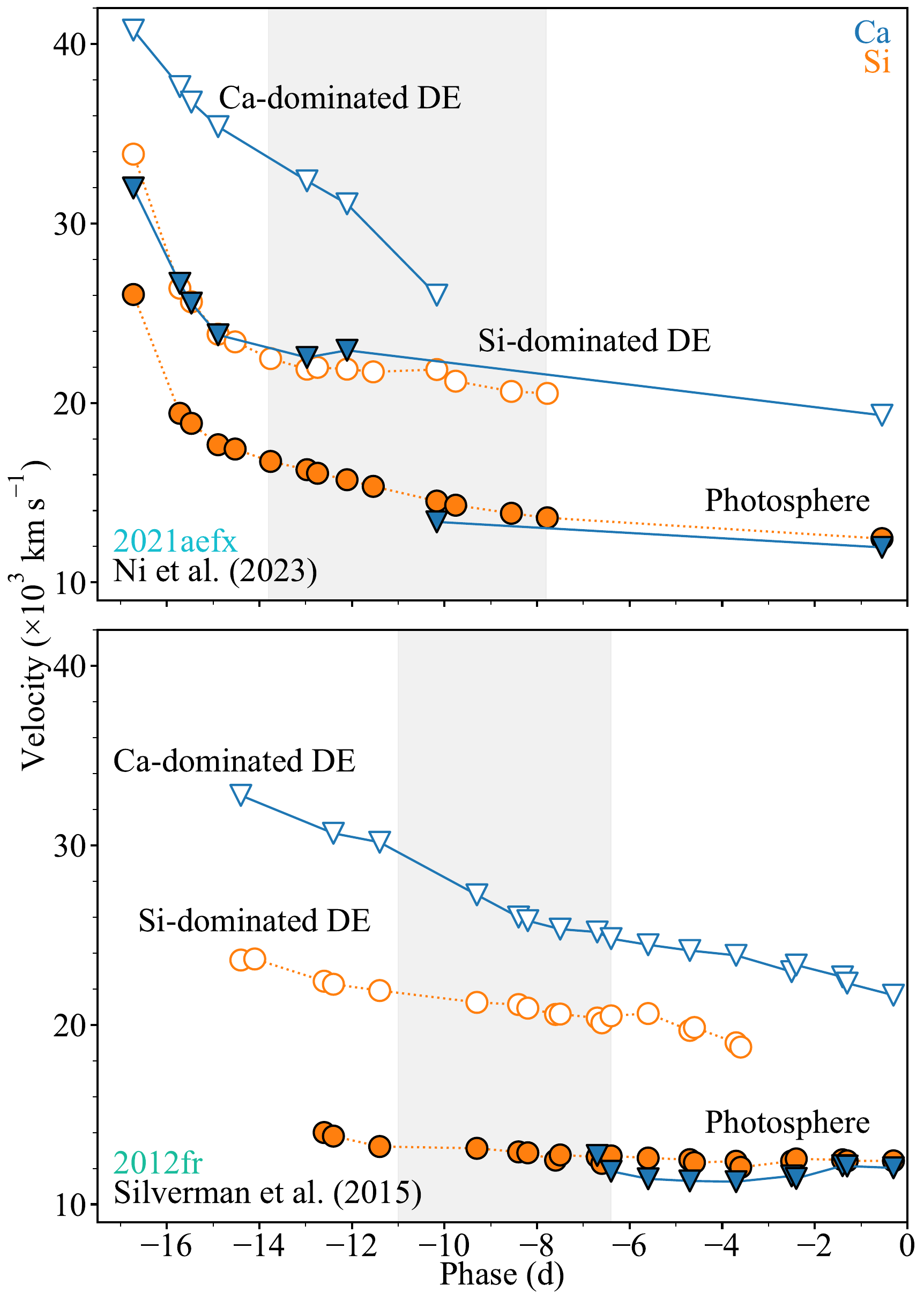}
 \caption{Velocity measurements of the PV and HV components of the \SiFeature\ (orange circles) and \CaII\ NIR (blue triangles) features in SN~2021aefx and SN~2012fr from \protect\cite{2021aefx_redshift_first_peak} and \protect\cite{2012fr_velocities}, respectively. The hollow points were components classified as HVFs in the literature with the solid points as those classified to be PVFs. The lines join together points that we propose to be forming in the same line forming regions, labelled as the photosphere, Si-dominated density enhancement (DE) and Ca-dominated DE. The phase ranges covered by the modelling in this paper are shaded in grey. The phases for the SN~2012fr measurements are given relative to the date of maximum used throughout this work and therefore, differ slightly from the values provided in \protect\cite{2012fr_velocities}.}
\label{fig:2021aefx_2012fr_velocities}
\end{figure}

If this is the case, we expect photospheric components of Ca and Si to follow very similar velocity evolutions, with the HVFs of the two species separated by several thousand \kms. Depending upon the Ca abundance in the vicinity of the Si-dominated density enhancement, we might see a third component of the \CaII\ NIR sitting between the typically seen PV and HV components, tracing the evolution of the \SiFeature\ HVF. This idea of these three line-forming regions is visualised in Fig.~\ref{fig:2021aefx_2012fr_velocities} where we plot the velocity measurements for the PV and HV components of the \SiFeature\ and \CaII\ NIR features of SN~2021aefx and SN~2012fr from \cite{2021aefx_redshift_first_peak} and \cite{2012fr_velocities} respectively. We use the velocities from these studies because they cover a broader phase range than our measurements, as well as having measurements of the \CaII\ NIR triplet. These literature measurements are in close agreement with the velocity measurements made in this study.

The \CaII\ NIR photospheric feature in SN~2021aefx identified by \cite{2021aefx_redshift_first_peak} (solid blue triangles) appears to closely trace the \SiFeature\ HVF up until $-12$~d at which point it drops by $\sim8000$~\kms\ in two days to then follow the evolution of the \SiFeature\ PVF. We instead propose the presence of three distinct \CaII\ NIR components: one at the photosphere which appears at $\sim-10$~d, a second formed by the Si-dominated density enhancement derived in the MCMC fitting above, and a third coming from a Ca-dominated density enhancement - corresponding to the HV component identified in \cite{2021aefx_redshift_first_peak}. The lack of evidence of a third silicon component implies a drop off in the silicon abundance at the highest velocities.

SN~2012fr similarly shows a late-forming photospheric \CaII\ NIR component that aligns in velocity space with the photospheric \SiFeature, and a HV \CaII\ NIR component several thousand kilometres per second faster than the \SiFeature\ HVF. However, \cite{2012fr_velocities} do not identify a \CaII\ NIR component that traces the evolution of the \SiFeature\ HVF, implying a lower calcium abundance for this object in the region of the Si-dominated density enhancement.

\begin{figure}
 \includegraphics[width = \columnwidth]{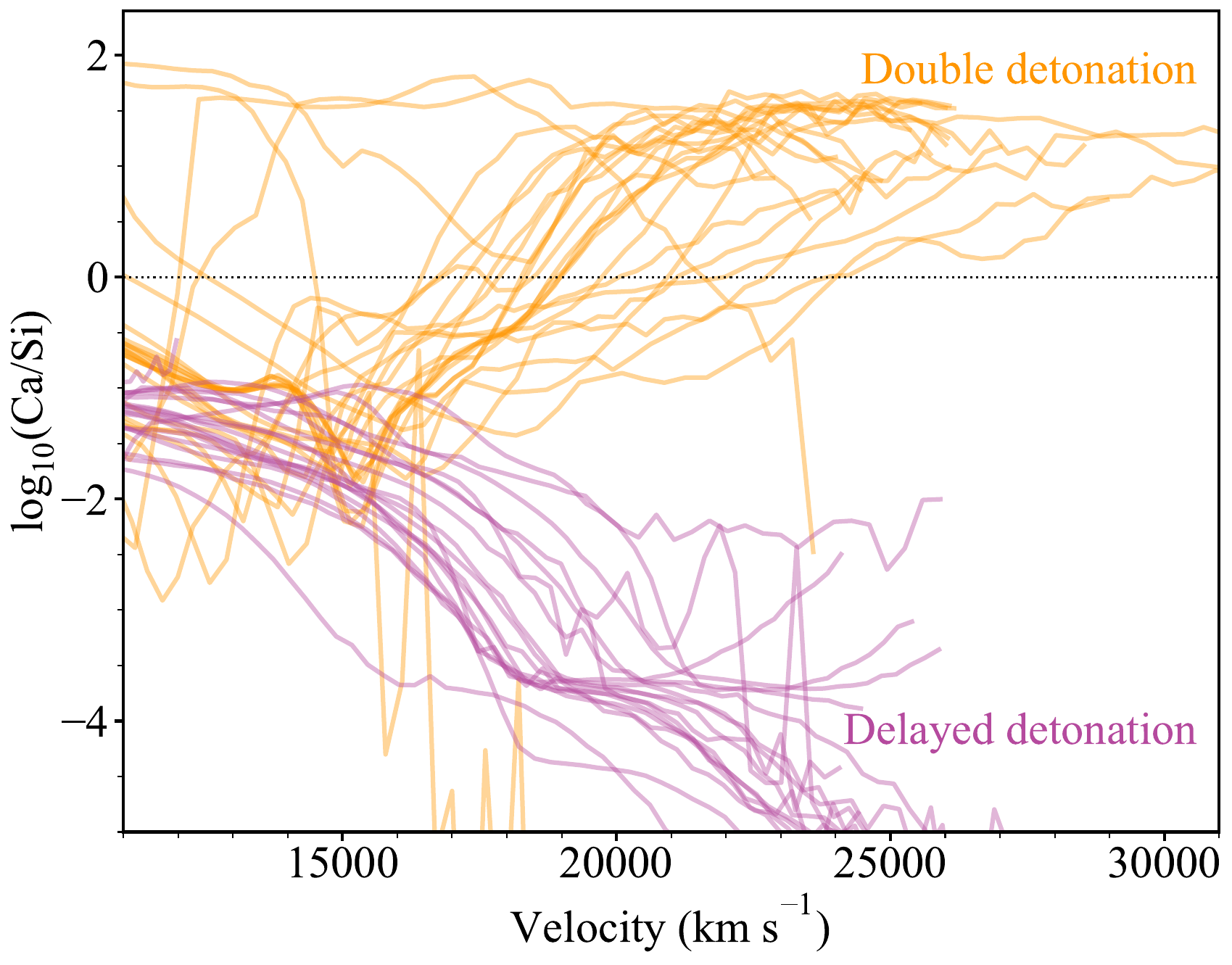}
 \caption{Ca/Si ratios as a function of velocity of the outer ejecta for the delayed-detonation explosion models of \cite{ddt_models} shown in pink and the double-detonation explosion models of \cite{double_det_models} shown in orange.}
\label{fig:CaSi_ratios}
\end{figure}

\subsection{Comparison to hydrodynamical models}
\label{sec:discussion:hydro}

Based on our argument in Section~\ref{sec:discussion:origin_HVSiII} that the density enhancements arise from the explosions themselves, and our hypothesis of three line-forming regions described in Section~\ref{sec:discussion:ca_HVFS}, we investigated if some common explosion models could provide plausible matches in the appropriate regions of the outer ejecta.

Delayed-detonations exhibit a single explosion site in the core, producing a high concentration of IGEs in the core transitioning through the IMEs to unburnt material moving out through the ejecta. The Ca abundance profile peaks at lower velocities than Si, however with significant overlap, resulting in similar PVF velocities for the two species. As seen however, the observed Ca HVFs are significantly faster than their Si counterparts, which is difficult to explain with the Ca peaking at lower velocities as in this model. The double-detonations instead have two explosion sites, in the core and in the shell. As such these models produce this IGE to IME gradient from the core outwards, and from the shell inwards, resulting in double-peaked abundance profiles, with the Ca outer peak sitting at higher velocities than that of Si. This behaviour can be seen in Fig.~\ref{fig:CaSi_ratios} where we plot the angle-averaged ratio of the abundances of Ca to Si for delayed-detonation and double-detonation explosion models from \cite{ddt_models} and \cite{double_det_models}, respectively. The two explosion mechanisms largely overlap at lower velocities -- however, they exhibit large separation as we move into the outer ejecta. The prediction drawn here of a low Ca/Si ratio increasing towards higher velocities generally matches the abundance ratios produced by the double-detonation models. The \SiFeature\ and \CaII\ NIR lines have different conditions for line formation, as such we emphasise that while these modelling results imply this increasing Ca/Si ratio, specific modelling of the HVFs of both features simultaneously would be required to constrain the amount of silicon that could be present in the region of the \CaII\ HVF and thus the range of the Ca/Si ratio.

\begin{figure}
 \includegraphics[width = \columnwidth]{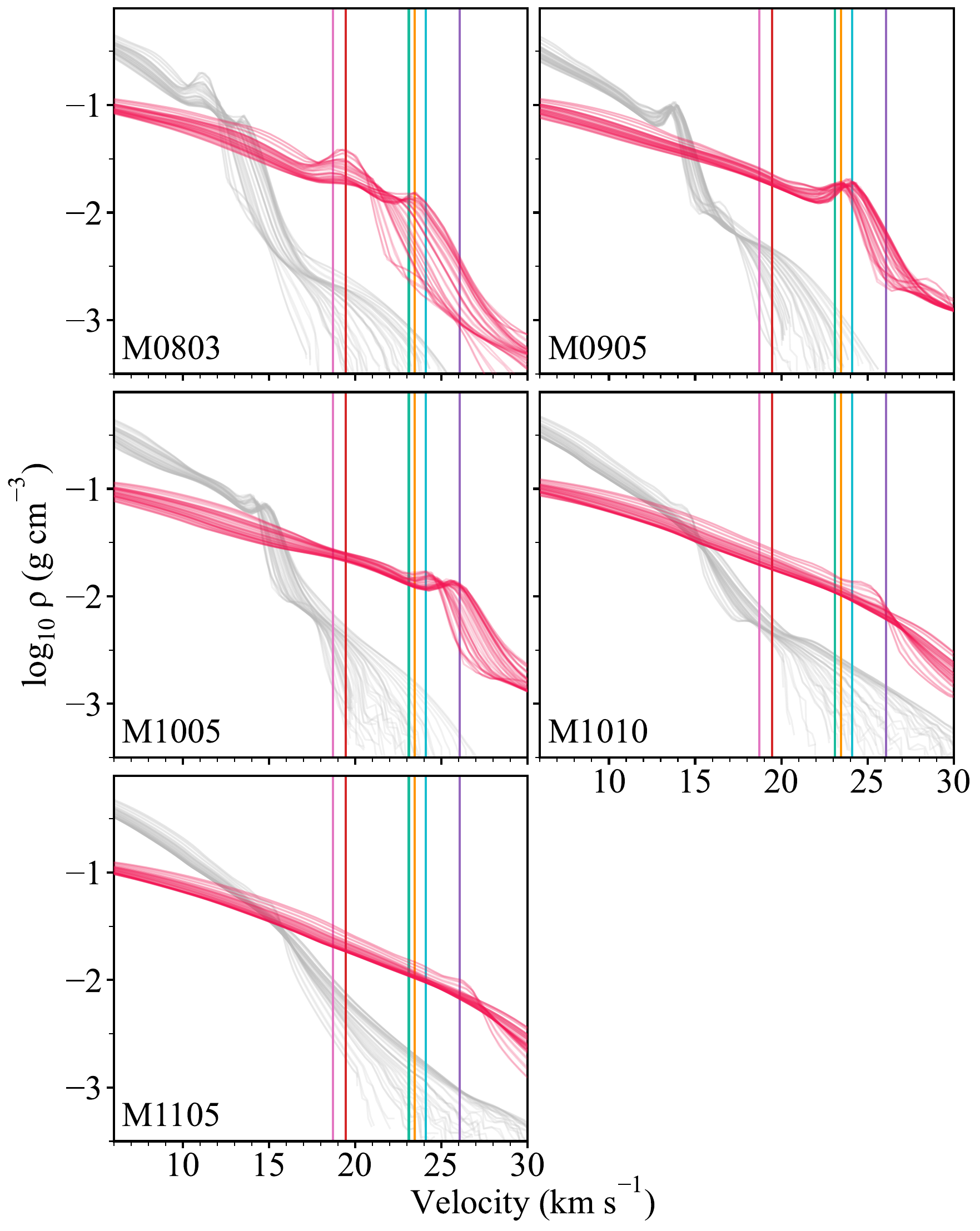}
 \caption{Density profiles from 100 lines of sight for five double-detonation models (grey). In pink are the same profiles scaled up to have 300\%\ the kinetic energy. The vertical coloured lines correspond to the peaks of the density enhancements ($b$ values) derived from the modelling for our six SNe, with the colours matching those for these objects used throughout this work.}
\label{fig:LOS_densities}
\end{figure}

From the intrinsic asymmetry of the double-detonation mechanism, we would expect variation in the density profiles from these models from various lines of sight. In Fig.~\ref{fig:LOS_densities} we present in grey the line-of-sight density profiles for 100 different lines of sight for five 3D double-detonation models (M0803, M0905, M1005, M1010 and M1105) from \cite{double_det_models} with varying core and shell masses. These models show density enhancements of varying strengths at velocities $\sim$10000 - 15000 \kms, with some variation seen between the lines of sign and between the models. However, these density enhancements lie at velocities far below those inferred for our six SNe from the modelling performed here of $\sim$19000 -- 26000 \kms. 

To investigate what would be required for a closer match with the data, we have scaled these density profiles to new kinetic energies using the following equations of \cite{Hachinger_ek_scaling} and \cite{Ashall_1986G}:
\begin{equation}
\label{eqn:density_scaling_1}
\centering
    \rho' = \rho_0\left(\frac{E'}{E_0}\right)^{-\frac{3}{2}}\left(\frac{M'}{M_0}\right)^{\frac{5}{2}}
\end{equation}
\begin{equation}
\label{eqn:density_scaling_2}
\centering
    v' = v_0\left(\frac{E'}{E_0}\right)^{\frac{1}{2}}\left(\frac{M'}{M_0}\right)^{-\frac{1}{2}}
\end{equation}
where $\rho$ is the density profile, $E$ is the kinetic energy, $v$ is the velocity profile, and $M$ is the mass (taken here to be constant). The subscript 0 denotes the initial values with the scaled values labelled with dashes. The pink lines in Fig.~\ref{fig:LOS_densities} display these line of sight density profiles with 300\%\ the kinetic energy of the base profiles. While the density enhancements from these scaled profiles align roughly with the locations of the inferred density enhancements from our modelling of each SNe Ia (vertical coloured lines in Fig.~\ref{fig:LOS_densities}), the requirement of a 300\%\ increase in kinetic energy is unrealistic and leads us to conclude that the double-detonation mechanism alone cannot explain the origin of HVFs.

Due to the more symmetric nature of the delayed-detonation mechanism, we find far less variation in the density profiles from different lines of sight, with the angle-averaged profiles representing well the general density profile. We do not see such clear density bumps in these delayed-detonation angle-averaged density profiles and as such the delayed-detonation mechanism alone is also insufficient to explain the origin of HVFs.

\subsection{The impacts of physics treatments}
\label{sec:discussion:limitations}
As outlined in Section~\ref{sec:method:modelling}, our simulations employ the \texttt{nebular} ionisation and \texttt{dilute-lte} excitation approximations, as implemented in \textsc{tardis}. These treatments approximate the complex physical processes governing ionisation and excitation in the SN ejecta, reducing computation times and enabling parameter space exploration. Since they introduce some limitations, it is important to assess whether they significantly affect our key conclusion: that a density enhancement in the outer ejecta is required to reproduce the observed HVFs.

\subsubsection{Excitation}
The \texttt{dilute-lte} excitation approximation assumes that level populations follow Boltzmann distributions at a local radiation temperature, scaled by a dilution factor, which is updated as part of the iterative MCMC fitting procedure. \cite{tardis_original} compared the \texttt{dilute-lte} treatment in \textsc{tardis} to full NLTE calculations by examining departure coefficients, which quantify deviations of level populations from LTE. They found that for the levels responsible for the \SiFeature\ feature, the difference between the departure coefficients from \texttt{dilute-lte} and the NLTE calculations was very small in the region around the photosphere, resulting in little change to the strength or shape of the formed line.

As we move higher into the ejecta and the densities drop, the approximation of LTE begins to break down as the radiation field decouples from the plasma. This is reflected by the marginal deviation of the departure coefficients from the two treatments several thousand \kms\ above the photosphere. While the level populations in the region of the photosphere are relatively unaffected by the choice of excitation treatment, this becomes slightly more important in the HVF region. The differences would primarily affect the precise shape and depth of the \SiFeature\ feature rather than eliminating it altogether. In particular, a full-NLTE excitation treatment may slightly alter the absorption strength or velocity width of the HVF, which would in turn shift the optimal values of the density enhancement parameters. These changes however are expected to be modest, and therefore the overall conclusion — that a density enhancement in the outer ejecta is required to reproduce the observed HVF — remains robust even under a more complete excitation treatment.

\subsubsection{Ionisation}
The \texttt{nebular} ionisation treatment is based on the modified nebular approximation of \cite{mazzali_lucy_1993}. It provides a computationally inexpensive approximation to NLTE ionisation without solving the full system of statistical equilibrium equations.

\begin{figure}
 \includegraphics[width = \columnwidth]{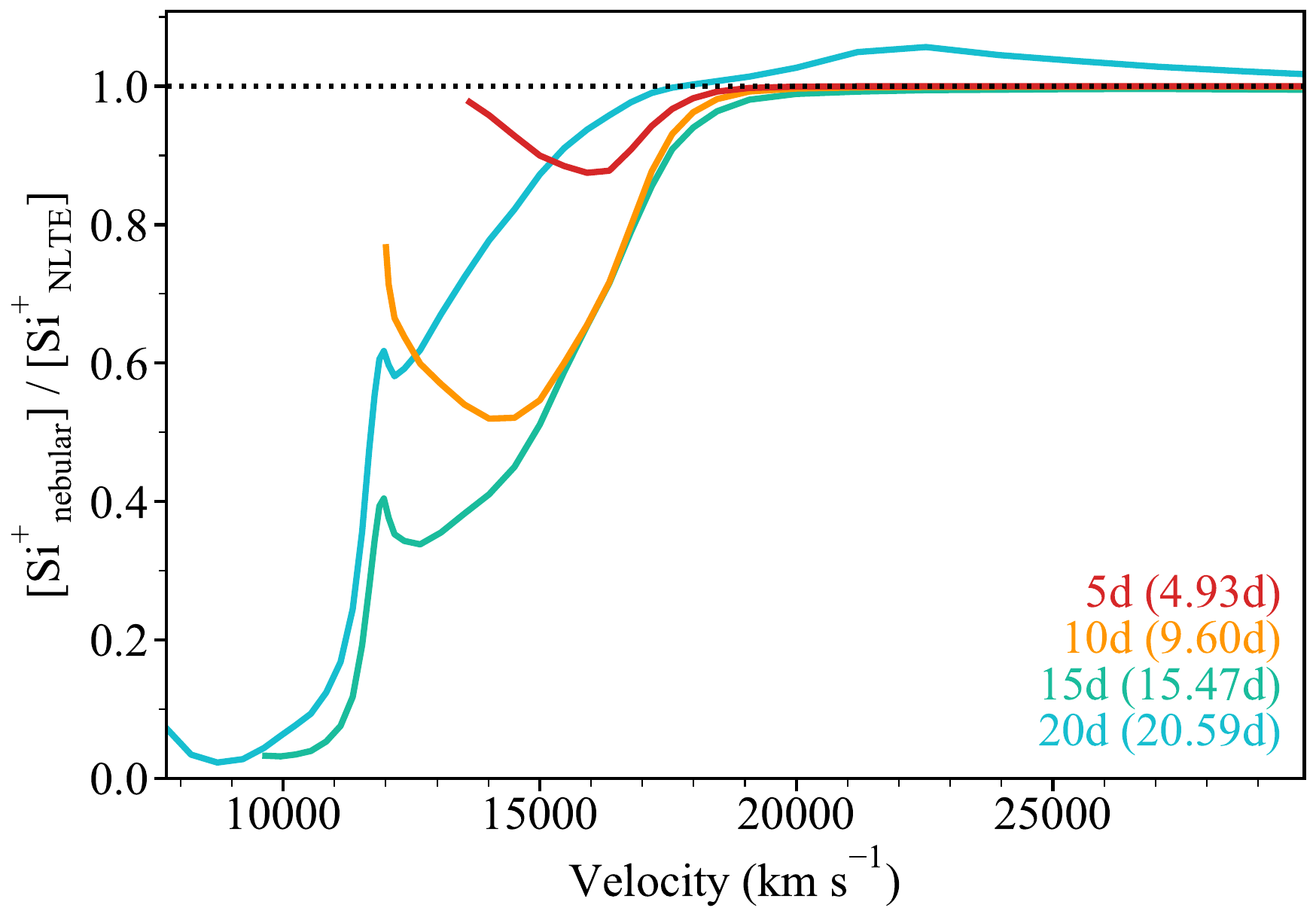}
 \caption{Ratio of the relative fraction of singly ionised silicon between the \textsc{tardis} and NLTE \textsc{cmfgen} simulations of the low-luminosity delayed-detonation model DDC25 \citep{ddc} as a function of velocity. Phases in brackets refer to the epochs of the \textsc{cmfgen} models, while those outside brackets correspond to the \textsc{tardis} times since explosion.}
\label{fig:ionisation_code_comparison}
\end{figure}

To assess its accuracy, we refer to the detailed radiative-transfer code comparison study \citep{standart_code_comparison}, which included \textsc{tardis} (using the \texttt{nebular} ionisation and \texttt{dilute-lte} excitation modes) and the 1D full-NLTE code \textsc{cmfgen} \citep{cmfgen1, cmfgen2}. Among other metrics, they compared the ionisation fractions of various elements at multiple epochs. In Fig.~\ref{fig:ionisation_code_comparison}, we present the ratio of singly ionised silicon predicted by \textsc{tardis} to that of \textsc{cmfgen} from a low-luminosity ($\sim0.1M_\odot$ of $^{56}$Ni) Chandrasekhar mass delayed-detonation model, which serves as a proxy for the deviation from full-NLTE ionisation. The times at which the \textsc{cmfgen} calculations were performed do not align perfectly with those of the \textsc{tardis} simulations, and we therefore take the closest epoch in each case. We see larger deviations from NLTE ionisation balance closer to the photosphere, with this discrepancy becoming more pronounced at later epochs. In the simulations up to 15~d post explosion, we find close agreement (within 1\%) at velocities above 20,000~\kms - the region of the density enhancements. This similarly holds true for the other low-luminosity model investigated.

The same ionisation balance comparison for higher-luminosity models -- $^{56}$Ni masses of $\sim0.6M_\odot$ -- yields however a different result. In these models we find that the \textsc{tardis} ionisation balance differs greatly from the NLTE calculations. This deviation is likely in part due to the \texttt{nebular} treatment, as at early phases the densities are many orders of magnitude above the regime of nebular theory and the ejecta is optically thick to the ionising UV photons. The photospheric approximation within \textsc{tardis} as the approximation of an opaque photosphere may also contribute to this deviation as it makes \textsc{tardis} agnostic to the presence of radioactive material deeper in the ejecta, which may affect the radiation field in ways not captured by the blackbody boundary condition.

We therefore caution that the \texttt{nebular} ionisation treatment may not be sufficiently accurate to determine whether density enhancements are required for the formation of HVFs. We strongly encourage specific modelling of \SiFeature\ and \CaII\ HVFs in full-NLTE to assess whether such enhancements are indeed necessary, or if NLTE effects alone can produce sufficient quantities of detached singly ionised material to reproduce the features observed here.

\section{Conclusion}
\label{conclusions}
Modelling of HVFs in the \SiFeature\ feature for six well studied literature SNe has been performed with the spectral-synthesis code \textsc{tardis} by way of introducing Gaussian density enhancements in the upper ejecta. A simplistic custom abundance profile was derived for each object through abundance tomography to match the PV features, with a high-velocity layer with uniform abundances. A grid of simulations with various density enhancements in the high velocity layer were then performed, and NNs trained upon the \SiFeature\ structure arising from the resulting grids. These networks were then used in an MCMC framework to infer the best fitting density enhancement for each target.

The extreme separations seen in the features of SN~2012fr were matched closely, with the quality of the fitting tending to decrease as the separation decreased and the features became more entangled. This is likely due to the fact that in cases of large velocity separation, the density enhancement sits much further from the photosphere and therefore the impact of the density enhancement upon the PV components and overall spectrum is smaller, making the base PV model more accurate. The SNe with larger velocity separations required stronger density enhancements, whereas the smaller separations were the result of weaker enhancements. We suggest that these extreme enhancements are harder to create as an explanation for the scarcity of HVFs with large separations from their PV counterparts.

By varying the Ca abundance in this outer layer of material we investigated the possibility of forming the Si and Ca HVFs from a single density enhancement. As expected the velocities of the resulting Ca HVFs are too low, leading us to speculate that there may be a secondary density enhancement higher up in the ejecta. This would make for three distinct line forming regions: the photosphere and then two density enhancements. We never see a third Si component formed by this second density enhancement, and only occasionally do we identify Ca components forming at the first density enhancement, suggesting that we require small Ca abundances compared to silicon at the photosphere which increases into the outer ejecta as the Si abundance drops - corresponding to an increasing Ca/Si ratio with increasing velocity. This pattern in Ca/Si can be seen in the models of the double-detonation mechanism however not in the delayed-detonation mechanism. Further investigation of this ratio would be interesting, with simultaneous modelling of \SiFeature\ and \CaII\ HVFs being required to place meaningful constraints.

While we find the density profiles of various double-detonation models from various lines of sight to produce a range of density enhancements, these occur at velocities far below those derived in this work. To match the velocities, the double-detonation models would require three times the kinetic energy and therefore, the double-detonation mechanism alone cannot explain the HVFs. Delayed detonations of Chandrasekhar mass models also fail to reproduce this high-velocity features. Further work is required both to obtain and model a larger sample of well-studied SNe Ia and to explore the outer regions of the ejecta in further multi-dimensional hydrodynamical explosion models. 

Currently there is no obviously favoured explosion mechanism that can describe the formation of HVFs. Based on the findings of this work, future spectroscopic modelling studies should aim to simultaneously reproduce the evolution of \SiII\ and \CaII\ HVFs. Such further investigation will allow us to constrain the absolute and relative abundances of these two key species, providing tighter constraints upon the composition of the outer ejecta. From an explosion modelling perspective, greater emphasis should be placed on understanding the high-velocity ejecta and the variation we see between different models and mechanisms in the high velocity regime. A comprehensive mechanism capable of explaining HVF formation is essential to account for the wide range of targets in which they are observed. Updated modelling including NLTE effects as detailed in Section \ref{sec:discussion:limitations} are also important to confirm our results. Furthermore, from a data collection standpoint, early-time spectra are critical for the rapid identification of HVFs, providing the necessary foundation for subsequent spectroscopic follow-up and detailed modelling.

\section*{Acknowledgements}
The research conducted in this publication was funded by the Irish Research Council under grant number GOIPG/2020/1387. K.M. acknowledges support from EU H2020 ERC grant no. 758638. K.M. and T.E.M.B. are supported by Horizon Europe grant no. 101125877. A.H. acknowledges support by the Klaus Tschira Foundation. A.H. is a Fellow of the International Max Planck Research School for Astronomy and Cosmic Physics at the University of Heidelberg (IMPRS-HD). J.P.A. acknowledges funding by ANID, Millennium Science Initiative, ICN12\_009. T.-W.C. acknowledges the Yushan Fellow Program by the Ministry of Education, Taiwan for the financial support (MOE-111-YSFMS-0008-001-P1).

Based on observations collected at the European Southern Observatory under ESO programme 106.216C.006/012.

This work made use of the Heidelberg Supernova Model Archive (HESMA), https://hesma.h-its.org.

\bibliographystyle{aa}

\appendix
\section{Figures}
\begin{figure}[h]
 \includegraphics[width = \columnwidth]{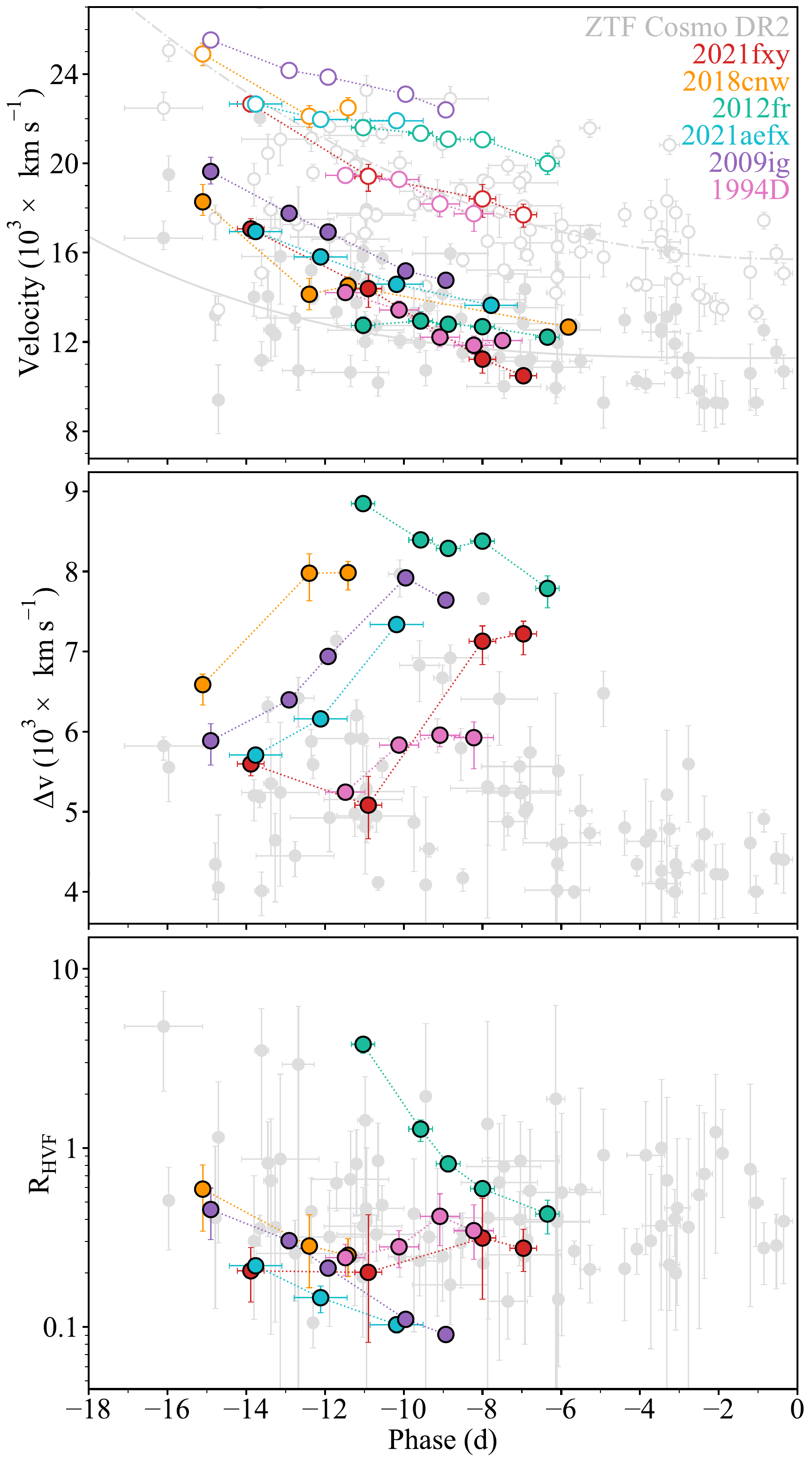}
 \caption{Comparison of the HVF SNe Ia sample collected here and the ZTF Cosmo DR2 HVF sample (grey) from H1. Here we plot the velocity evolution of the PV (solid) and HV (hollow) components (top), the separation between them in velocity space (middle), and the ratio of their pEWs (bottom). The same code has been used to fit both samples. The colours for each object match those in Fig.~\ref{fig:spectral_series} and used throughout. The solid and dot-dashed lines in the velocity panel correspond to the fits to the PV and HV components from the DR2 sample (H1), respectively.}
 \label{fig:dr2_comparison}
\end{figure}

\begin{figure}
 \includegraphics[width = \columnwidth]{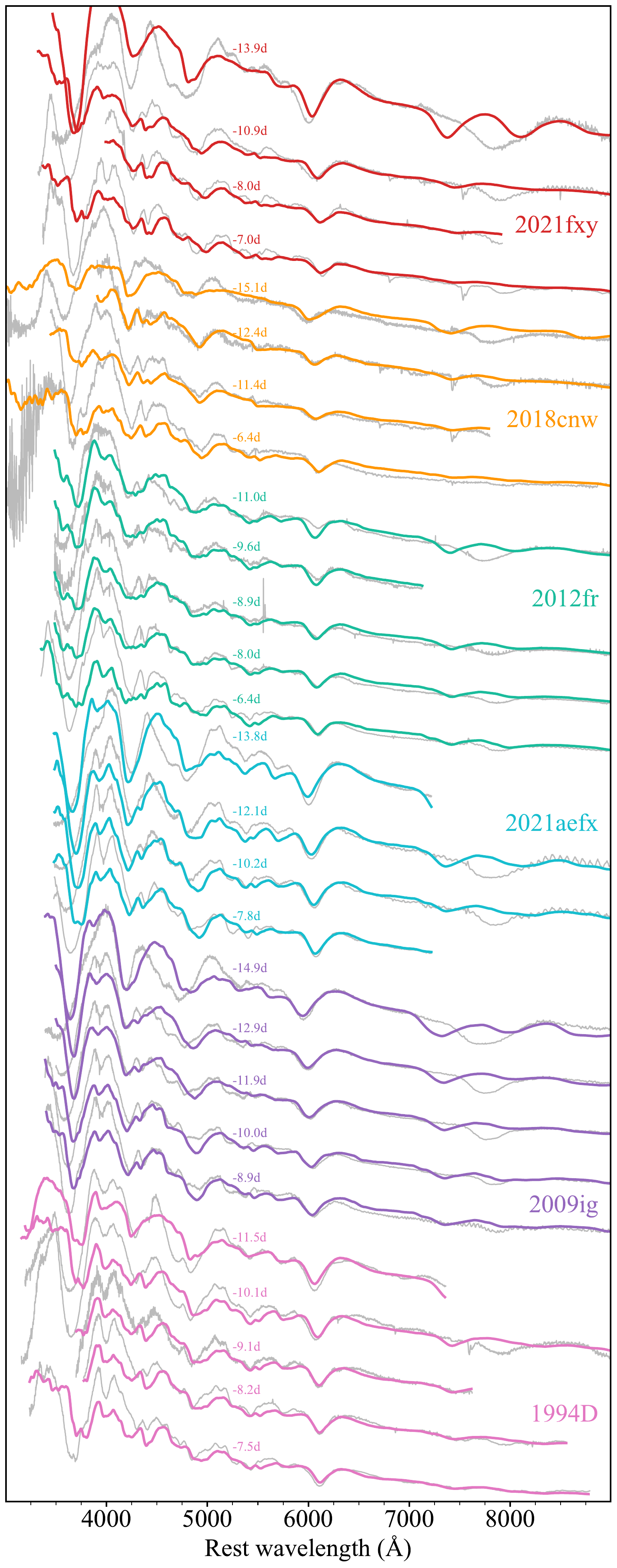}
 \caption{Photospheric feature models for our six SNe Ia. The synthetic model spectra are shown in colour for each object with the observed spectra from Fig.~\ref{fig:spectral_series} in grey. To aid comparison between the PV models and the observed \SiFeature\ PV components, the HV components have been removed from the observed spectra by adding the best fitting HV component Gaussian fit to the spectrum.}
 \label{fig:PV_spectral_series}
\end{figure}

\begin{figure*}
 \includegraphics[width = \linewidth]{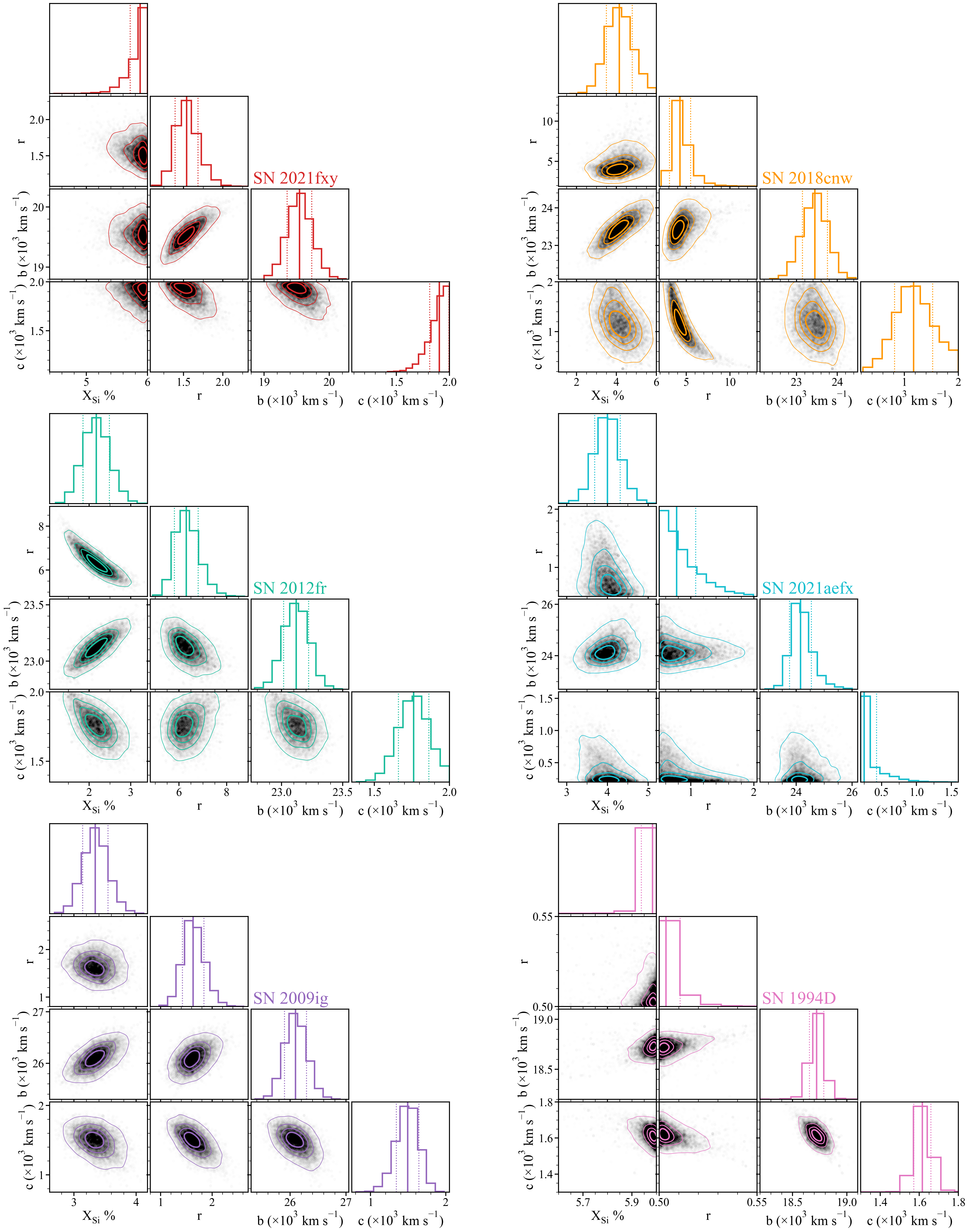}
 \caption{Corner plots of the parameter posteriors inferred by the density enhancement MCMC runs for the 6 SNe. The grid boundaries are indicated as grey shaded regions in panels where they are visible. The solid and dotted lines in the histogram panels correspond to the means and standard deviations respectively.}
 \label{fig:corner_plots}
\end{figure*}

\onecolumn
\section{Tables}
\begin{table}[h]
\caption{Spectroscopic data of the 27 epochs used in our study.}
\label{tab:spectral_log}
\centering
\begin{tabular}{lccccc|ccc}
\hline
\textbf{Target} & \textbf{MJD} & \textbf{Phase (d)} & \textbf{Telescope} & \textbf{Instrument} & \textbf{Source} & \textbf{t$_{\text{exp}}$} (d) & \textbf{L} ($\log$ L$_\odot$) & \textbf{v$_{\text{inner}}$} (km s$^{-1}$)\\ \hline
2021fxy & 59291.11 & $-13.9\pm0.3$ & SALT & RSS & \textit{a} & 4.5 & 8.44 & 15300 \\
2021fxy & 59294.17 & $-10.9\pm0.3$ & NTT & EFOSC2 & this paper & 7.53 & 9.13 & 12400 \\
2021fxy & 59297.04 & $-8.0\pm0.3$ & LT & SPRAT & this paper & 10.38 & 9.43 & 11900 \\
2021fxy & 59298.09 & $-7.0\pm0.3$ & NOT & ALFOSC & \textit{b} & 11.42 & 9.49 & 11600 \\
2018cnw & 58286.55 & $-15.1\pm0.1$ & Keck & LRIS & \textit{c}  & 5.5 & 8.8 & 18000 \\
2018cnw & 58289.33 & $-12.4\pm0.1$ & APO & DIS & \textit{c} & 8.2 & 9.32 & 13900 \\
2018cnw & 58290.34 & $-11.4\pm0.1$ & DCT & Deveny+LMI & \textit{c} & 9.2 & 9.42 & 13800 \\
2018cnw & 58296.06 & $-5.8\pm0.1$ & NOT & ALFOSC & \textit{c} & 14.8 & 9.65 & 12200 \\
2012fr & 56231.53 & $-11.0\pm0.3$ & ANU-2.3m & WiFeS & \textit{d} & 8.04 & 9.15 & 13100 \\
2012fr & 56232.99 & $-9.6\pm0.3$ & SAAO-1.9m & GS & \textit{d} & 9.5 & 9.31 & 13000 \\
2012fr & 56233.69 & $-8.9\pm0.3$ & ANU-2.3m & WiFeS & \textit{d} & 10.19 & 9.35 & 12975 \\
2012fr & 56234.57 & $-8.0\pm0.3$ & ANU-2.3m & WiFeS & \textit{d} & 11.07 & 9.42 & 12800 \\
2012fr & 56236.22 & $-6.4\pm0.3$ & NTT & EFOSC2 & \textit{d} & 12.71 & 9.49 & 12090 \\
2021aefx & 59532.84 & $-13.8\pm0.7$ & SALT & RSS & \textit{e} & 4.5 & 8.51 & 16000 \\
2021aefx & 59534.50 & $-12.1\pm0.7$ & FTS & FLOYDS & \textit{e} & 6.15 & 8.88 & 14300 \\
2021aefx & 59536.44 & $-10.2\pm0.7$ & FTS & FLOYDS & \textit{e} & 8.08 & 9.18 & 13900 \\
2021aefx & 59538.85 & $-7.8\pm0.7$ & SALT & RSS & \textit{e} & 10.48 & 9.40 & 13600 \\
2009ig & 55065.51 & $-14.9\pm0.1$ & Lick-3m & KAST & \textit{f} & 3.5 & 8.47 & 18125 \\
2009ig & 55067.52 & $-12.9\pm0.1$ & Lick-3m & KAST & \textit{f} & 5.5 & 8.95 & 16400 \\
2009ig & 55068.52 & $-11.9\pm0.1$ & Lick-3m & KAST & \textit{f} & 6.5 & 9.1 & 16000 \\
2009ig & 55070.50 & $-10.0\pm0.1$ & Lick-3m & KAST & \textit{f} & 8.5 & 9.27 & 14800 \\
2009ig & 55071.53 & $-8.9\pm0.1$ & Lick-3m & KAST & \textit{f} & 9.5 & 9.34 & 14600 \\
1994D & 49420.00 & $-11.5\pm0.5$ & Lick-3m & KAST & \textit{g} & 5.5 & 8.65 & 14000 \\
1994D & 49421.36 & $-10.1\pm0.5$ & MMT & MMTBlue & \textit{h} & 6.86 & 8.97 & 12200 \\
1994D & 49422.41 & $-9.1\pm0.5$ & FLWO-1.5m & FAST & \textit{h} & 7.91 & 9.14 & 12100 \\
1994D & 49423.27 & $-8.2\pm0.5$ & ESO-3.6m & EFOSC1 & \textit{i} & 8.77 & 9.26 & 12000 \\
1994D & 49424.00 & $-7.5\pm0.5$ & Keck & LRIS & \textit{g} & 9.5 & 9.34 & 11800 \\ \hline
\end{tabular}
\begin{flushleft}
    \textbf{Notes.} Further details on the telescopes and instruments can be found in the relevant listed source. The listed phases are with respect to maximum light in the rest-frame of the supernova. The final three columns present the \textsc{tardis} PV model inputs for each of the modelled spectra: rest-frame time since explosion (t$_{\text{exp}}$), requested luminosity (L), and inner boundary velocity (v$_\text{inner}$). \\
    \textbf{References.} \textit{a}: \protect\cite{2021fxy_spectra1}, obtained from the Transient Name Server\protect{\footnote{https://www.wis-tns.org}}, \textit{b}: \protect\cite{2021fxy_spectra2}, \textit{c}: \protect\cite{Rigault_DR2}, \textit{d}: \protect\cite{2012fr_spectra1}, \textit{e}: \protect\cite{2021aefx_spectra1}, \textit{f}: \protect\cite{2009ig_spectra1}, \textit{g}: \protect\cite{1994D_spectra1}, \textit{h}: \protect\cite{1994D_spectra2}, and \textit{i}: \protect\cite{1994D_spectra3}.
\end{flushleft}
\end{table}

\begin{table}[h]
\caption{Photometry data table for the six SNe Ia studied.}
\label{appendix:tab:photometry}
\begin{tabular}{lccccccc}
\hline
\textbf{Target} & \textbf{\textit{z}} & \textbf{$\mu$ (mag)} & \textbf{MW $A_v$} & \textbf{First light (MJD)} & \textbf{Peak (MJD)} & \textbf{Flux calibration bands} \\ \hline
2021fxy & 0.0094$^{a}$ & 32.87(0.09)$^{b}$ & 0.25327(0.0028) & 59289.71(0.16) & 59305.12(0.34)$^{b}$ & \textit{g r} (SWOPE)$^{b}$ \\
2018cnw & 0.024167$^{d}$ & 35.07$^{*}$ & 0.07161(0.0031) & 58283.04(0.28) & 58302.02(0.05)$^{d}$ & \textit{g r} (ZTF)$^{d}$ \\
2012fr & 0.005457$^{e}$ & 31.27(0.05)$^{f}$ & 0.05487(0.0012) & 56225.73(0.46)$^{g}$ & 56242.6(0.3)$^{g}$ & \textit{g r} (SWOPE)$^{g}$ \\
2021aefx & 0.005284$^{h}$ & 31.27(0.49)$^{i}$ & 0.02449(0.00062) & 59529.85(0.55)$^{h}$ & 59546.67(0.67)$^{h}$ & \textit{g r} (SWOPE)$^{j}$ \\
2009ig & 0.008770$^{k}$ & 32.60(0.4)$^{l}$ & 0.08339(0.0025) & 55063.41(0.08)$^{m}$ & 55080.54(0.04)$^{m}$ & \textit{B R} (KAIT4)$^{m}$ \\
1994D & 0.002058$^{n}$ & 31.02(0.19)$^{o}$ & 0.05921(0.00031) & - & 49431.5(0.5)$^{p}$ & \textit{B R}$^{p}$ \\\hline
\end{tabular}
\begin{flushleft}
    \textbf{Notes.} Uncertainties are noted in parenthesis for values calculated in this work and when given for values taken from the literature. Milky Way extinction corrections were performed using $R_v=3.1$, with \textit{E(B-V)} values through querying the IRSA dust maps \protect\citep{IRSA_dustmap}.\\
$^{*}$Calculated from the redshift measurement with H$_0=70.3$~km~s$^{-1}$~Mpc \protect{\citep{hubble_constant}}.\\
    \textbf{References.} \textit{a}: \protect\cite{2021fxy_redshift}, \textit{b}: \protect\cite{2021fxy_spectra2}, \textit{c}: \protect\cite{2018cnw_redshift}, \textit{d}: \protect\cite{Rigault_DR2}, \textit{e}: \protect\cite{2012fr_redshift}, \textit{f}: \protect\cite{2012fr_distance_modulus}, \textit{g}: \protect\cite{2012fr_dates_photometry}, \textit{h}: \protect\cite{2021aefx_redshift_first_peak}, \textit{i}: \protect\cite{2021aefx_distance_modulus}, \textit{j}: \protect\cite{2021aefx_photometry}, \textit{k}: \protect\cite{2009ig_redshift}, \textit{l}: \protect\cite{2009ig_distance_modulus}, \textit{m}: \protect\cite{2009ig_spectra1}, \textit{n}: \protect\cite{1994D_redshift}, \textit{o}: \protect\cite{1994D_distance_modulus}, \textit{p}: \protect\cite{1994D_peak_photometry}. 
\end{flushleft}
\end{table}
\twocolumn

\begin{landscape}
\begin{table}
\caption{Fit parameters to the observed silicon absorption features.}
\label{tab:silicon_hvf_fit_parameters}
\begin{tabular}{lccccccccccc}
\hline\textbf{ZTF Name} & \textbf{Phase (d)} & \textbf{a$_{\text{PV}}$} & \textbf{b$_{\text{PV}}$ (\r A)} & \textbf{c$_{\text{PV}}$ (\r A)} & \textbf{a$_{\text{HV}}$} & \textbf{b$_{\text{HV}}$ (\r A)} & \textbf{c$_{\text{HV}}$ (\r A)} & \textbf{pEW$_{\text{PV}}$ (\r A)} & \textbf{pEW$_{\text{HV}}$ (\r A)} & \textbf{R} & \textbf{$\Delta v$ (km s$^{-1}$)} \\ \hline
1994D & $-11.5\pm{0.5}$ & $0.287\pm^{0.004}_{0.004}$ & $6060.5\pm^{2.6}_{3.2}$ & $79.8\pm^{2.0}_{1.6}$ & $0.129\pm^{0.011}_{0.012}$ & $5955.0\pm^{1.6}_{1.6}$ & $42.9\pm^{1.7}_{1.8}$ & $115.0\pm^{4.3}_{3.7}$ & $27.8\pm^{3.3}_{3.6}$ & $0.24\pm^{0.04}_{0.05}$ & $5244\pm^{65}_{95}$ \\ [1.55pt]
1994D & $-10.1\pm{0.5}$ & $0.272\pm^{0.005}_{0.004}$ & $6076.3\pm^{3.3}_{3.9}$ & $76.4\pm^{2.5}_{2.1}$ & $0.126\pm^{0.012}_{0.013}$ & $5958.6\pm^{3.2}_{2.8}$ & $45.8\pm^{2.2}_{2.3}$ & $104.5\pm^{4.8}_{4.5}$ & $29.0\pm^{4.2}_{4.3}$ & $0.28\pm^{0.06}_{0.07}$ & $5833\pm^{47}_{72}$ \\ [1.55pt]
1994D & $-9.1\pm{0.5}$ & $0.244\pm^{0.009}_{0.011}$ & $6101.2\pm^{4.7}_{5.7}$ & $68.4\pm^{3.0}_{2.5}$ & $0.119\pm^{0.015}_{0.016}$ & $5980.6\pm^{7.7}_{7.0}$ & $56.6\pm^{3.4}_{3.4}$ & $83.6\pm^{6.4}_{6.5}$ & $34.0\pm^{6.4}_{6.5}$ & $0.42\pm^{0.14}_{0.13}$ & $5956\pm^{87}_{146}$ \\ [1.55pt]
1994D & $-8.2\pm{0.5}$ & $0.233\pm^{0.007}_{0.012}$ & $6108.6\pm^{4.1}_{4.2}$ & $61.9\pm^{2.1}_{1.9}$ & $0.084\pm^{0.012}_{0.010}$ & $5989.4\pm^{11.7}_{8.9}$ & $57.8\pm^{5.0}_{4.2}$ & $72.3\pm^{4.7}_{5.6}$ & $24.4\pm^{5.8}_{4.7}$ & $0.35\pm^{0.14}_{0.11}$ & $5925\pm^{196}_{388}$ \\ [1.55pt]
1994D & $-7.5\pm{0.5}$ & $0.240\pm^{0.001}_{0.001}$ & $6104.3\pm^{0.3}_{0.3}$ & $70.7\pm^{0.5}_{0.5}$ & - & - & - & $85.0\pm^{0.6}_{0.6}$ & - & - & - \\ [1.55pt]
2009ig & $-14.9\pm{0.0}$ & $0.251\pm^{0.011}_{0.011}$ & $5951.6\pm^{7.1}_{8.6}$ & $91.9\pm^{4.3}_{3.5}$ & $0.179\pm^{0.023}_{0.025}$ & $5835.1\pm^{2.8}_{2.4}$ & $57.1\pm^{2.3}_{2.7}$ & $115.7\pm^{10.3}_{9.0}$ & $51.4\pm^{8.8}_{9.5}$ & $0.45\pm^{0.14}_{0.15}$ & $5887\pm^{213}_{302}$ \\ [1.55pt]
2009ig & $-12.9\pm{0.0}$ & $0.224\pm^{0.002}_{0.002}$ & $5989.0\pm^{1.6}_{2.1}$ & $91.8\pm^{1.4}_{1.0}$ & $0.130\pm^{0.004}_{0.005}$ & $5861.9\pm^{0.9}_{0.9}$ & $47.8\pm^{0.9}_{1.0}$ & $103.2\pm^{2.1}_{1.8}$ & $31.2\pm^{1.6}_{1.8}$ & $0.30\pm^{0.03}_{0.03}$ & $6397\pm^{48}_{68}$ \\ [1.55pt]
2009ig & $-11.9\pm{0.0}$ & $0.207\pm^{0.001}_{0.001}$ & $6005.9\pm^{1.0}_{1.4}$ & $94.1\pm^{1.1}_{0.9}$ & $0.097\pm^{0.002}_{0.003}$ & $5867.8\pm^{0.7}_{0.8}$ & $42.6\pm^{0.8}_{0.9}$ & $97.8\pm^{1.3}_{1.2}$ & $20.8\pm^{0.9}_{1.0}$ & $0.21\pm^{0.01}_{0.02}$ & $6939\pm^{32}_{47}$ \\ [1.55pt]
2009ig & $-10.0\pm{0.0}$ & $0.193\pm^{0.001}_{0.001}$ & $6041.0\pm^{0.3}_{0.4}$ & $81.2\pm^{0.5}_{0.3}$ & $0.056\pm^{0.001}_{0.001}$ & $5882.9\pm^{0.7}_{0.7}$ & $30.7\pm^{0.5}_{0.5}$ & $78.6\pm^{0.4}_{0.4}$ & $8.7\pm^{0.2}_{0.2}$ & $0.11\pm^{0.00}_{0.00}$ & $7920\pm^{23}_{32}$ \\ [1.55pt]
2009ig & $-8.9\pm{0.0}$ & $0.214\pm^{0.001}_{0.001}$ & $6049.4\pm^{0.5}_{0.6}$ & $78.4\pm^{0.7}_{0.6}$ & $0.050\pm^{0.000}_{0.000}$ & $5896.8\pm^{1.7}_{1.6}$ & $30.2\pm^{0.4}_{0.2}$ & $84.1\pm^{0.8}_{0.8}$ & $7.6\pm^{0.1}_{0.1}$ & $0.09\pm^{0.00}_{0.00}$ & $7643\pm^{57}_{94}$ \\ [1.55pt]
2012fr & $-11.0\pm{0.3}$ & $0.092\pm^{0.004}_{0.003}$ & $6090.4\pm^{2.4}_{3.5}$ & $56.3\pm^{2.5}_{1.9}$ & $0.236\pm^{0.001}_{0.002}$ & $5912.6\pm^{1.4}_{1.6}$ & $83.4\pm^{1.4}_{1.6}$ & $26.1\pm^{1.9}_{1.4}$ & $98.6\pm^{1.8}_{2.2}$ & $3.79\pm^{0.33}_{0.42}$ & $8848\pm^{66}_{100}$ \\ [1.55pt]
2012fr & $-9.6\pm{0.3}$ & $0.127\pm^{0.003}_{0.003}$ & $6086.5\pm^{3.2}_{4.5}$ & $61.6\pm^{3.5}_{2.6}$ & $0.154\pm^{0.003}_{0.003}$ & $5917.5\pm^{3.1}_{3.2}$ & $64.3\pm^{2.8}_{2.7}$ & $39.4\pm^{2.6}_{2.2}$ & $49.8\pm^{2.4}_{2.8}$ & $1.27\pm^{0.16}_{0.18}$ & $8394\pm^{70}_{97}$ \\ [1.55pt]
2012fr & $-8.9\pm{0.3}$ & $0.148\pm^{0.002}_{0.002}$ & $6089.4\pm^{1.5}_{2.2}$ & $64.1\pm^{2.5}_{1.7}$ & $0.131\pm^{0.002}_{0.002}$ & $5922.8\pm^{2.1}_{2.4}$ & $58.9\pm^{1.7}_{1.8}$ & $47.7\pm^{2.2}_{1.7}$ & $38.7\pm^{1.4}_{1.7}$ & $0.82\pm^{0.07}_{0.08}$ & $8288\pm^{49}_{69}$ \\ [1.55pt]
2012fr & $-8.0\pm{0.3}$ & $0.162\pm^{0.001}_{0.001}$ & $6091.6\pm^{0.5}_{0.6}$ & $62.9\pm^{0.8}_{0.6}$ & $0.114\pm^{0.001}_{0.001}$ & $5923.2\pm^{0.7}_{0.8}$ & $53.1\pm^{0.6}_{0.7}$ & $51.1\pm^{0.7}_{0.5}$ & $30.3\pm^{0.4}_{0.5}$ & $0.59\pm^{0.02}_{0.02}$ & $8379\pm^{20}_{29}$ \\ [1.55pt]
2012fr & $-6.4\pm{0.3}$ & $0.187\pm^{0.003}_{0.003}$ & $6101.1\pm^{1.9}_{2.4}$ & $55.1\pm^{2.5}_{1.9}$ & $0.073\pm^{0.003}_{0.004}$ & $5944.4\pm^{6.0}_{5.2}$ & $59.5\pm^{5.9}_{6.0}$ & $51.8\pm^{2.6}_{2.3}$ & $21.9\pm^{2.9}_{3.2}$ & $0.43\pm^{0.08}_{0.10}$ & $7788\pm^{156}_{242}$ \\ [1.55pt]
2021aefx & $-13.8\pm{0.7}$ & $0.300\pm^{0.001}_{0.001}$ & $6005.4\pm^{0.9}_{1.1}$ & $85.2\pm^{0.7}_{0.5}$ & $0.139\pm^{0.003}_{0.004}$ & $5891.6\pm^{0.5}_{0.6}$ & $40.4\pm^{0.7}_{0.7}$ & $128.3\pm^{1.5}_{1.3}$ & $28.1\pm^{1.1}_{1.3}$ & $0.22\pm^{0.01}_{0.02}$ & $5707\pm^{24}_{37}$ \\ [1.55pt]
2021aefx & $-12.1\pm{0.7}$ & $0.297\pm^{0.002}_{0.002}$ & $6028.3\pm^{1.8}_{2.1}$ & $82.7\pm^{1.8}_{1.4}$ & $0.104\pm^{0.007}_{0.008}$ & $5905.2\pm^{1.9}_{2.0}$ & $34.1\pm^{2.0}_{2.0}$ & $123.5\pm^{2.9}_{2.5}$ & $17.8\pm^{2.1}_{2.1}$ & $0.15\pm^{0.02}_{0.03}$ & $6160\pm^{55}_{81}$ \\ [1.55pt]
2021aefx & $-10.2\pm{0.7}$ & $0.278\pm^{0.002}_{0.002}$ & $6053.1\pm^{0.8}_{1.0}$ & $76.9\pm^{1.1}_{0.8}$ & $0.069\pm^{0.003}_{0.004}$ & $5906.4\pm^{2.1}_{2.1}$ & $31.6\pm^{1.6}_{1.2}$ & $107.3\pm^{1.4}_{1.2}$ & $11.0\pm^{0.8}_{0.8}$ & $0.10\pm^{0.01}_{0.01}$ & $7338\pm^{69}_{95}$ \\ [1.55pt]
2021aefx & $-7.8\pm{0.7}$ & $0.275\pm^{0.001}_{0.001}$ & $6072.2\pm^{0.3}_{0.3}$ & $71.1\pm^{0.4}_{0.4}$ & - & - & - & $98.0\pm^{0.6}_{0.5}$ & - & - & - \\ [1.55pt]
2021fxy & $-13.9\pm{0.3}$ & $0.273\pm^{0.007}_{0.007}$ & $6003.1\pm^{4.4}_{5.1}$ & $91.1\pm^{2.6}_{2.3}$ & $0.094\pm^{0.016}_{0.016}$ & $5891.6\pm^{3.1}_{2.8}$ & $53.5\pm^{3.1}_{3.5}$ & $125.0\pm^{6.4}_{6.3}$ & $25.2\pm^{6.0}_{5.8}$ & $0.21\pm^{0.07}_{0.07}$ & $5597\pm^{97}_{149}$ \\ [1.55pt]
2021fxy & $-10.9\pm{0.3}$ & $0.204\pm^{0.012}_{0.019}$ & $6057.0\pm^{13.1}_{9.3}$ & $91.5\pm^{5.3}_{6.0}$ & $0.073\pm^{0.034}_{0.019}$ & $5955.8\pm^{9.4}_{6.8}$ & $47.5\pm^{10.4}_{9.6}$ & $94.4\pm^{8.9}_{13.5}$ & $17.5\pm^{12.9}_{7.1}$ & $0.20\pm^{0.22}_{0.12}$ & $5083\pm^{359}_{422}$ \\ [1.55pt]
2021fxy & $-8.0\pm{0.3}$ & $0.172\pm^{0.007}_{0.007}$ & $6121.3\pm^{8.6}_{8.3}$ & $72.4\pm^{7.9}_{6.9}$ & $0.083\pm^{0.015}_{0.015}$ & $5976.1\pm^{12.4}_{8.7}$ & $44.5\pm^{10.4}_{8.0}$ & $62.9\pm^{7.3}_{7.4}$ & $18.5\pm^{7.9}_{6.1}$ & $0.31\pm^{0.21}_{0.17}$ & $7129\pm^{191}_{289}$ \\ [1.55pt]
2021fxy & $-7.0\pm{0.3}$ & $0.190\pm^{0.003}_{0.003}$ & $6136.5\pm^{2.3}_{2.5}$ & $60.3\pm^{2.2}_{1.8}$ & $0.059\pm^{0.004}_{0.004}$ & $5990.4\pm^{7.2}_{5.8}$ & $52.2\pm^{6.9}_{5.8}$ & $57.7\pm^{2.3}_{2.3}$ & $15.6\pm^{3.1}_{2.7}$ & $0.28\pm^{0.08}_{0.07}$ & $7221\pm^{158}_{263}$ \\ [1.55pt]
2018cnw & $-15.1\pm{0.1}$ & $0.160\pm^{0.008}_{0.007}$ & $5978.7\pm^{8.1}_{11.5}$ & $77.9\pm^{7.7}_{5.2}$ & $0.130\pm^{0.016}_{0.022}$ & $5847.4\pm^{6.3}_{5.7}$ & $55.1\pm^{3.2}_{3.8}$ & $62.7\pm^{9.1}_{6.6}$ & $36.1\pm^{6.5}_{8.4}$ & $0.59\pm^{0.22}_{0.25}$ & $6587\pm^{130}_{252}$ \\ [1.55pt]
2018cnw & $-12.4\pm{0.1}$ & $0.126\pm^{0.005}_{0.005}$ & $6062.1\pm^{10.1}_{10.5}$ & $111.1\pm^{8.1}_{7.3}$ & $0.065\pm^{0.013}_{0.012}$ & $5902.4\pm^{6.2}_{5.3}$ & $59.1\pm^{6.6}_{6.8}$ & $70.6\pm^{7.2}_{6.9}$ & $19.4\pm^{6.0}_{5.1}$ & $0.28\pm^{0.14}_{0.12}$ & $7979\pm^{244}_{343}$ \\ [1.55pt]
2018cnw & $-11.4\pm{0.1}$ & $0.125\pm^{0.003}_{0.003}$ & $6054.6\pm^{3.2}_{4.1}$ & $78.4\pm^{4.5}_{3.9}$ & $0.055\pm^{0.004}_{0.004}$ & $5894.9\pm^{5.5}_{5.2}$ & $44.0\pm^{4.5}_{4.5}$ & $49.4\pm^{3.1}_{2.8}$ & $12.3\pm^{1.8}_{1.8}$ & $0.25\pm^{0.06}_{0.06}$ & $7985\pm^{143}_{216}$ \\ [1.55pt]
2018cnw & $-5.8\pm{0.1}$ & $0.189\pm^{0.002}_{0.002}$ & $6092.0\pm^{1.0}_{0.9}$ & $57.6\pm^{1.1}_{1.1}$ & - & - & - & $54.5\pm^{1.1}_{1.1}$ & - & - & - \\ [1.55pt] \hline
\end{tabular}
\end{table}
\end{landscape}

\end{document}